\documentclass[fleqn,usenatbib]{mnras}
\pdfoutput=1
\usepackage{newtxtext,newtxmath}

\usepackage[T1]{fontenc}
\usepackage{ae,aecompl}

\usepackage{graphicx}	
\usepackage{amsmath}	
\usepackage{amssymb}	
\usepackage{bm}
\usepackage{booktabs}
\usepackage{mathrsfs}
\graphicspath{{Figures/}}
\usepackage{ulem}
\usepackage[dvipsnames]{xcolor}

\newcommand{\xmark}{\text{X}}

\title[A pitfall of piecewise-polytropic equation of state inference]{A pitfall of piecewise-polytropic equation of state inference}

\author[G. Raaijmakers et al.]{
Geert Raaijmakers,\thanks{E-mail: G.Raaijmakers@uva.nl}
Thomas E. Riley,
Anna L. Watts
\\
Anton Pannekoek Institute for Astronomy, University of Amsterdam, PO Box 94249, 1090GE Amsterdam, the Netherlands
}

\date{Accepted XXX. Received YYY; in original form ZZZ}

\pubyear{2018}

\begin{document}
\label{firstpage}
\pagerange{\pageref{firstpage}--\pageref{lastpage}}
\maketitle

\begin{abstract}
The only messenger radiation in the Universe which one can use to statistically probe the Equation of State (EOS) of cold dense matter is that originating from the near-field vicinities of compact stars. Constraining gravitational masses and equatorial radii of rotating compact stars is a major goal for current and future telescope missions, with a primary purpose of constraining the EOS. From a Bayesian perspective it is necessary to carefully discuss prior definition; in this context a complicating issue is that in practice there exist pathologies in the general relativistic mapping between spaces of local (interior source matter) and global (exterior spacetime) parameters. In a companion paper, these issues were raised on a theoretical basis. In this study we reproduce a probability transformation procedure from the literature in order to map a joint posterior distribution of Schwarzschild gravitational masses and radii into a joint posterior distribution of EOS parameters. We demonstrate computationally that EOS parameter inferences are sensitive to the choice to define a prior on a joint space of these masses and radii, instead of on a joint space interior source matter parameters. We focus on the piecewise-polytropic EOS model, which is currently standard in the field of astrophysical dense matter study. We discuss the implications of this issue for the field.
\end{abstract}

\section{Introduction}
\label{sec:introduction}

Neutron star cores reach supranuclear densities: higher than those which can be probed by Earth-based experiments. Matter in cores, if nucleonic, is very neutron rich. However, at these extreme densities phase transitions to non-nucleonic matter may also occur: hybrid stars may form, whose cores consist of deconfined quarks \citep{Collins75}; hyperons may form \citep{Ambartsumyan60}; or a colour superconducting phase may manifest \citep{Alford98, Rapp98, Alford99}.
Some theories even predict stars entirely made out of deconfined up, down, and strange quarks \citep[see, e.g.,][]{Haensel86} or stars which contain a Bose-Einstein condensate of pions \citep{Sawyer72, Sawyer73} or kaons \citep{Kaplan86}. For recent comprehensive reviews of possible states of matter in the neutron star core, see \citet{Oertel17} and \citet{Baym18}.
The uncertain microphysics of the particle interactions are often described by the equation of state (EOS) of cold, dense matter; a simple relation between pressure and energy density. In general relativistic gravity the local pressure and energy density appear in the Einstein field equations: it is possible, therefore, to probe thermodynamical properties -- i.e., the EOS -- of astrophysical cold dense matter as a way to understanding microphysics at supranuclear matter densities.

The only way to access the part of the EOS corresponding to the density regime in neutron star interiors is by statistically quantifying how well thermodynamical (and microphysical) models describe observable reality. There are many models with fixed microphysical parameters \citep[for examples, see][]{Lattimer01}. Increasingly however, studies are using parametrised models, of which the most common are piecewise-polytropic models \citep{Read09, Ozel10, Hebeler10, Steiner10, Steiner13, Raithel16}; an alternative involves spectral representation of the EOS \citep{Lindblom12}. These models can be tested by using astronomical data sets (e.g., X-ray data); such testing requires application of the general relativistic stellar structure equations to map the dense matter EOS to parameters appearing in analytical exterior spacetime solutions, such as gravitational mass, equatorial radius, and moment of inertia. Hereafter we refer to these parameters as exterior parameters. For static neutron stars the structure equations are the Tolman-Oppenheimer-Volkoff equations \citep{Tolman39, Oppenheimer39}, while for slowly rotating neutron stars these equations have to be modified using the Hartle-Thorne formalism to include rotation \citep*{Hartle67, Hartle68}. For rapidly rotating neutron stars one can make use of exact numerical solutions of the structure equations with open source codes such as \texttt{RNS} \citep{Stergioulas1995}. 
With this mapping, mass, radius and moment of inertia can be used to constrain the dense matter EOS \citep[for recent papers on the topic see][]{Steiner10, Hebeler13, Steiner15, Ozel16, Raithel16a, Raithel17}.

There exist several observational techniques for statistically estimating these exterior spacetime parameters with reasonable accuracies. In recent years, the most stringent mass constraints have been inferred by modelling the orbital motion of binary pulsars which are radio timing targets \citep{Lattimer12}. There exist particularly strong constraints conditioned on observations of the two massive pulsars PSR J$0348+0432$ \citep{Antoniadis13} and PSR J$1614-2230$ \citep{Demorest10}, with reported masses of $2.01 \pm 0.04$ and $1.97 \pm 0.04$ M$_{\odot}$ respectively \citep[the latter revised to $1.93 \pm 0.02$ M$_{\odot}$ by][]{Fonseca16}. 

Accurately constraining radii has proven more difficult. Most current estimates make use of spectral modelling of X-ray emission from quiescent LMXBs or from thermonuclear bursts \citep[see, e.g.,][and references therein]{Miller16}. An alternative technique is that of X-ray pulse-profile modelling (also known as light-curve or waveform modelling), which exploits (exterior) spacetime effects on rotationally pulsed emission from surface emission anisotropies \citep[see, e.g.,][for a general introduction to the technique]{Watts16}. These X-ray techniques for estimating radii are being exploited by the recently-launched telescope NICER \citep{NICER}, and would also be used by proposed large area X-ray telescope concepts such as STROBE-X \citep{STROBEX} and eXTP \citep{eXTP}.

In this paper we focus on transforming (joint) statistical constraints on the masses and radii of non-rotating stars into constraints on parameters of an EOS model. Nevertheless, we note that \citet{Lattimer05} and \citet{Kramer09} expect that the moment of inertia of pulsar A in the double pulsar system PSR J$0737-3039$ will be estimated to a precision of $10 \%$ \citep{Lattimer05, Kramer09} within the next decade. In principle, moments of inertia can be used in the same manner as masses and radii to infer the EOS. Further, binary tidal deformation of neutron stars, inferred from gravitational wave observations in the final stages of inspiral, can also potentially constrain the cold EOS \citep{Lackey15, Abbott17}. These studies also aim to use parametrised models of the EOS.  

Calculation of the EOS, given \textit{known} values of Schwarzschild gravitational masses and radii, was first demonstrated by \citet{Lindblom92}. A non-parametric, numerical method was developed for the inversion of a discrete mass-radius relationship to an EOS; \citet{Lindblom92} acknowledged, however, that it would not be realistic with a limited number of known mass-radius pairs. The \citet{Lindblom92} approach is non-statistical: a parametric form of the EOS is necessary to make statistical inference numerically tractable.

Some studies using a parametric EOS model have focused on constraining the EOS using neutron star mass constraints alone: it is typically required that each candidate EOS permits a mass-radius relation that reaches the highest inferred mass \citep{Read09, Steiner10, Hebeler13,Kurkela14}.  The statistical approach of \citet[hereafter OP09]{Ozel10} can be interpreted from a Bayesian perspective: a joint posterior distribution of masses and radii is transformed onto the space of EOS parameters. OP09 treat the specific case of estimating three EOS parameters from three observed neutron stars, under the ansatz that the joint posterior distribution of the masses and radii of the neutron stars is a product of three bivariate Gaussians,\footnote{Note that each two-dimensional Gaussian distribution is defined on a distinct parameter space, such that the joint posterior distribution is six-dimensional.} each with standard deviations of $5 \%$ of the modal mass and radius. The authors argue that given a $5 \%$ standard deviation on each mass and radius, the posterior distribution of EOS parameters is sufficiently \textit{informative} for one to distinguish between three thermodynamically distinct EOSs.

However, Riley et al. (submitted; hereafter R18) argued that a piecewise-polytropic parametrisation does not admit an invertible mapping between interior source matter parameters (including EOS parameters) and exterior spacetime parameters. Therefore, if one conditions on the piecewise-polytropic model and invokes the \textit{Exterior-Prior} (EP) paradigm described in R18 (and based on OP09), the prior distribution on the space of the interior parameters will exhibit undesirable properties such as being informative \citep[e.g.,][]{Robert2007,Gelman_book},\footnote{Which from a more objectivist perspective is concerning \citep[e.g.,][and references therein]{Robert2007}.} or even ill-behaved. The posterior distribution (and any derived parameter inferences) may thus be described as \textit{distorted} by an ill-defined prior. Here we define the distortion as being relative to the distribution obtained when invoking the R18 \textit{Interior-Prior} (IP) paradigm: direct EOS parameter inference given astronomical data sets, with some well-behaved -- and optionally noninformative \citep[e.g.,][]{Robert2007} -- prior distribution defined on the space of interior parameters with physically-motivated bounds (see Section~\ref{sec:EOS parameter space definition}). See Section \ref{sec:parameterest} for a detailed explanation of this issue.

We begin in Section \ref{sec:methods} by describing the methods used for EOS parameter estimation; we also explain the workings of our code. In Section \ref{sec:results1} we explore the effect of conditioning on a piecewise-polytropic parametrisation, whilst changing the joint posterior distribution of masses and radii relative to the distribution considered by OP09. We discuss the implications of our results in Section \ref{sec:discussion}. 

\section{Methods}
\label{sec:methods}
\subsection{Statistical Framework}
\subsubsection{The EOS parametrisation}
\label{sec:polytropes}
We begin this section by describing the parametrisation of the EOS used throughout the paper. 
If the true, assumedly universal dense matter EOS is to exist within the model space, one may need an infinite number of basis functions and thus coefficients (parameters). Computational parameter estimation would then be intractable. It is thus necessary to use a parametric functional form of the EOS with a small number of parameters. 

The most commonly used parametrisation in the literature is a piecewise-polytropic model \citep{Mueller85}, which makes use of the sensitivity (in general relativistic gravity) of the mass and radius of a set of neutron stars to pressure at just a few fiducial transition densities, as shown by \citet{Read09} \citep[see also][]{Lattimer01}. Their parametrisation uses two transition densities at $\rho_{t,1} = 1.85 \rho_{ns}$ and $\rho_{t,2} = 2 \rho_{t,1}$, and a piecewise-polytropic relation between them, with the polytropic indices defined as free parameters. The authors also show that a large number of candidate EOSs, calculated from theoretical microphysics, can be reproduced with such a model to about $4 \%$ accuracy. A similar parametrisation was used by OP09, but instead of the polytropic indices, the pressures at three transition densities are defined as free parameters, with the third transition density being $\rho_{t,3} = 2 \rho_{t,2}$. The full parametrisation is completely described by three free parameters, $\bm{\theta} = (P_1,P_2,P_3)$, and can be written as
\begin{equation}
P(\rho) = K_i \rho^{\Gamma_i}, \ \ \ \ \rho_{i-1} \leq \rho \leq \rho_i 
\label{eqn:polytrope ranges}
\end{equation}
for $i = 1, 2, 3$, with the polytropic indices defined by the pressures at the transition densities
\begin{equation}
\Gamma_i = \frac{\log_{10}(P_i/P_{i-1})}{\log_{10}(\rho_{t,i}/\rho_{t,i-1})}.
\end{equation}
At $\rho_{t,0}=10^{14.3}$ g/cm$^3$ the parametrised EOS is connected to the SLy EOS \citep{Douchin01} such that the EOS is continuous at zeroth-order.

For a central density to be considered physically reasonable for a given EOS, we impose the fundamental condition that the adiabatic speed of sound never exceeds the speed of light for densities below that central density \citep[for a discussion on the validity of this condition, see, e.g.,][]{Ellis07}.
Analytically we write the condition of causality as
\begin{equation}
\label{causal}
\frac{d P}{d\varepsilon} \equiv \left(\frac{c_s}{c}\right)^2 \leq 1.
\end{equation}

Recently, \citet{Raithel16} found that using five polytropes with logarithmically spaced transition densities can significantly increase the accuracy with which theoretical candidate EOSs can be reproduced. However, using five polytropes (instead of three) increases the computational expense of parameter estimation due to a greater number of parameter dimensions; further, because the distortions we aim to demonstrate are independent of the number of polytropes, three polytropes are sufficient in this work.

\subsubsection{Equation of state parameter estimation}
\label{sec:parameterest}
To calculate the joint probability distribution of EOS parameters given a joint probability distribution of exterior spacetime parameters, we use the framework described in Section 2.3.2 of R18. In this framework, EOS parameter estimation is distilled into two phases: in the first phase a (Bayesian) joint posterior distribution of exterior parameters is calculated conditional on some astronomical data set; in the second phase this joint posterior distribution is transformed onto the space of interior parameters. 

Let us define, in analogy with R18 (appendix A6), a multivariate random variable $\bm{z}$  as a vector of exterior parameters. We then define the multivariate random variable $\boldsymbol{y}=(\boldsymbol{\theta},\boldsymbol{\rho_c})$, a vector of interior parameters where $\boldsymbol{\theta}$ consists of $n$ EOS parameters and $\boldsymbol{\rho_c}$ consists of central densities for $s$ stars. Let us define the mapping\footnote{Note that for, e.g., static stars, the mapping $h$ is given by the Tolman-Oppenheimer-Volkoff equations.} between interior and exterior parameters as $h\colon Y\to Z$, $\boldsymbol{y}\mapsto\bm{z}$, where $Y\subset\mathbb{R}^{n+s}$ is the domain of $h$, and $Z\subset\mathbb{R}^{d}$ is the codomain of map $h$ (of which the image of $Y$ under $h$ is a subset; see R18). The vectors of interior and exterior parameters are then elements of $Y$ and $Z$ respectively. In order to meaningfully transform a joint probability density distribution between the continuous sets $Z$ and $Y$ (respectively defined on the spaces $\mathbb{R}^{d}$ and $\mathbb{R}^{n+s}$), the map $h$ needs to be \textit{invertible}. 

In order for a map to be invertible, a number of conditions need to be satisfied (see R18, section 2.3.3, 3.1, 3.2 and Appendix A3). In particular, as the vectors $\bm{z}$ and $\boldsymbol{y}$ consist only of continuous parameters, the dimensionalities of $\bm{z}$ and $\boldsymbol{y}$ must be equal. Furthermore, the Jacobian determinant 
\begin{equation}
\label{eqn:jacobian}
J(\boldsymbol{y}) = \left| \det\left(\frac{\partial \bm{z}}{\partial \boldsymbol{y}}\right) \right|
\end{equation}
must be everywhere finite.  
If we only consider masses and radii, the dimensionality of $\bm{z}$ is $d = 2 s$. The dimensionality of $\boldsymbol{y}$ is the sum of $n$ and $s$. The requirement for the dimensionalities to be equal then reduces to $2s=n + s$ such that $s = n$. 
Given that the Jacobian determinant is locally well-defined, a local joint probability density of the parameters $\boldsymbol{y}$ can be simply related to a local joint probability density of the parameters $\bm{z}$, where $\bm{z} = h(\bm{y})$. More specifically, the joint posterior distribution of the EOS parameters is obtained by taking the product of the posterior distribution of masses and radii and the (modulus of the) Jacobian determinant. The central densities of stars are not of particular interest, and are therefore marginalised over to obtain the final distribution of the EOS parameters.

Analytically, the joint posterior distribution of the EOS parameters, $\mathcal{P}(\bm{\theta} \,|\, \mathcal{D}, \mathcal{M}, \mathcal{I})$, conditional on the entire data set $\mathcal{D}=\cup_{i}\mathcal{D}_{i}$, the global model $\mathcal{M}$, and prior information (independent data sets) $\mathcal{I}$, can be written as
\begin{equation}
\label{PosteriorRho}
\begin{aligned}
\mathcal{P}(\bm{\theta} ~|~ \mathcal{D}, \mathcal{M}, \mathcal{I}) \propto & \mathop{\int} \left[\prod_{i=1}^s \mathcal{P} (M_i, R_i ~|~ \mathcal{D}_{i}, \mathcal{M}, \mathcal{I})\right]J(\bm{y})d\bm{\rho}_{c}, \\
\end{aligned}
\end{equation}
where the distributions $\mathcal{P} (M_i, R_i \,|\, \mathcal{D}_{i}, \mathcal{M}, \mathcal{I})$ are normalised (marginal) joint posteriors of masses and radii, and $J(\bm{y})$ is defined by Equation (\ref{eqn:jacobian}). Note that the upper limit in each dimension of this integral is the maximum central density, $\rho_{c,\textrm{max}}(\boldsymbol{\theta})$, for which Equation~(\ref{causal}) still holds given an EOS $\boldsymbol{\theta}$. This is the non-rotating limit of equation (33) in section 2.5 of R18. We require a proportionality instead of an equality because the mapping $h$ is not injective (one-to-one) nor surjective (onto); a distributional renormalisation is required because the integral given by Equation~(\ref{PosteriorRho}) -- a function of the EOS parameters -- is not normalised on the EOS parameter space. For a full discussion refer to R18.

\begin{figure*}
  \includegraphics[width=\textwidth]{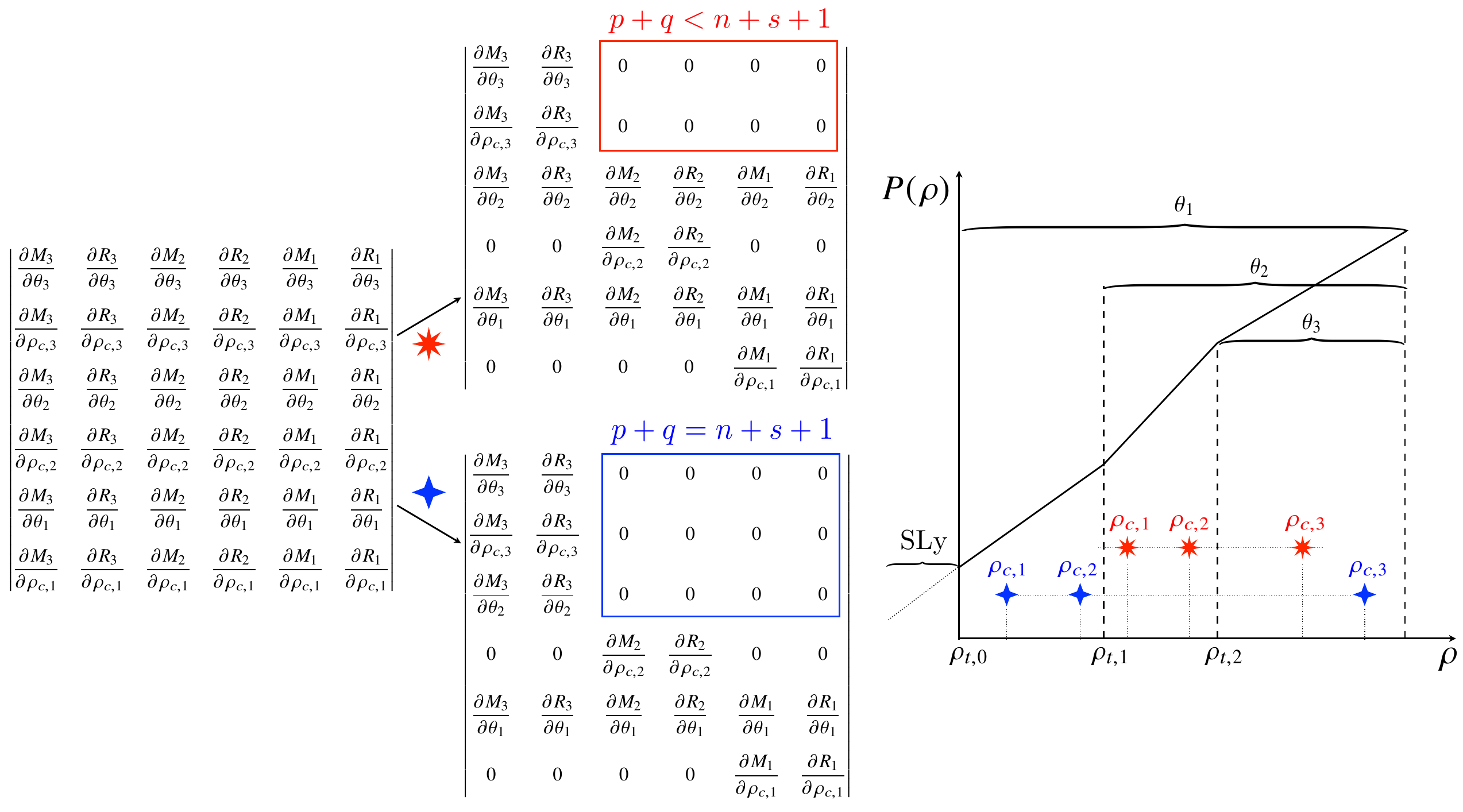}
  \caption{An illustration of the Jacobian determinant at two points $\bm{y}\in Y$, where we consider central densities of three stars. In the \textit{rightmost} panel we schematically depict the piecewise-polytrope EOS model for some parameter vector $\boldsymbol{\theta}$; the density subdomain spanned by an EOS parameter $\theta_{i}$ denotes the region in which the central density of a model star can lie such that a fraction of matter in the interior is thermodynamically dependent on $\theta_{i}$ (see Section \ref{sec:jacobian}). The \textit{leftmost} determinant is the general form with some ordering of rows and columns which permits faster visual inspection of whether the Jacobian is singular or not; we employ the condition that for the largest Jacobian submatrix of zeros of size $p\times q$, if $p+q\geq n+s+1$, the determinant is zero. The \textit{topmost} Jacobian matrix corresponds to a configuration illustrated by the red stars in the \textit{rightmost} panel, from which it is clear that the condition given in Equation~(\ref{eqn:non-singular condition}) is satisfied; the matrix is non-singular, as can be verified by noting that $p+q<n+s+1$. It follows that this Jacobian is associated with local invertibility at a point $\boldsymbol{y}$. The \textit{bottommost} Jacobian, on the other hand, does \textit{not} satisfy the condition given in Equation~(\ref{eqn:non-singular condition}), as illustrated by the blue stars in the \textit{rightmost} panel; the matrix is thus singular, as can be verified by evaluating that $p+q=n+s+1$.}
 \label{fig:jacdeterminant}
\end{figure*}

When invoking the EP-paradigm, the joint posterior distribution of EOS parameters can be distorted relative to the distribution of EOS parameters which would be calculated under the R18 IP-paradigm with, e.g., a minimally informative joint prior distribution \citep[e.g.,][]{Robert2007} defined on the space of EOS parameters with physically-motivated bounds. We will now briefly explain the origin of the distortion; for an in-depth treatment, see R18.

Suppose via the IP-paradigm that a point $\bm{y}$ has finite posterior support conditional on an informative astronomical data set. The likelihood of $\bm{y}$, $\mathcal{L}(\bm{y})$, is defined as the probability density of the data given $\bm{y}$, and because $\bm{y}$ and $\bm{z}$ are deterministically related through the map $h$, the equality $\mathcal{L}(\bm{y}) \equiv \mathcal{L}(\bm{z})$ must be true. On the other hand, suppose that one adopts an objectivist Bayesian viewpoint and defines a minimally informative prior (conventionally termed \textit{noninformative}) on the space of exterior parameters $\bm{z}$. Such a prior will not be invariant to exterior-interior reparametrisation because the mapping is non-invertible. If, e.g., the Jacobian is singular at the point $\boldsymbol{y}\mapsto\boldsymbol{z}$: (i) the joint prior probability density at point $\bm{z}$ cannot be transformed to a joint prior probability density at point $\bm{y}$; (ii) if an ill-defined transformation is performed whereby the determinant is applied to the prior density at point $\bm{z}$ irrespective of the singularity, the prior density at point $\bm{y}$ is, undesirably, zero. It follows that the joint prior distribution of interior parameters $\bm{y}$, and as a result the joint posterior distribution of $\bm{y}$, is distorted relative to the distribution calculated via direct estimation of interior parameters. The posterior distortion cannot be objectively quantified because the prior cannot strictly be locally transformed under $\boldsymbol{y}\mapsto{\boldsymbol{z}}$; the above working definition of \textit{distortion} is thus adopted throughout this paper.

We now focus on the piecewise-polytropic model, which is widely used in the literature to model the EOS in neutron star cores. As explained in R18 (section 2.3.4), when the EOS model is piecewise-polytropic (see Section \ref{sec:polytropes}), the Jacobian determinant is \textit{not} everywhere well-defined: each parameter $\theta_i$ in $\bm{\theta}$ only controls the EOS above  $\rho_{t,i}$, the associated transition density. If a neutron star has a central density lower than this transition density, the mass and radius of the star are not influenced by the part of the EOS that is controlled by $\theta_i$.  In other words, there is no matter in the star at densities above the transition density. As a result, the partial derivatives (at all orders) of $M$ and $R$ with respect to $\theta_i$ locally vanish. If this is the case for all $s$ stars, the Jacobian determinant is zero at point $\bm{y}$. For the Jacobian to be non-singular at $\bm{y}$ the following condition must be satisfied $\forall i \in 1\ldots n$:
\begin{equation}
 \rho_{c, i} \geq \rho_{t,i-1},
 \label{eqn:non-singular condition}
\end{equation}
where the central densities $\rho_{c,i}$ are such that central density increases with $i$. We illustrate this condition in Fig. \ref{fig:jacdeterminant} for two scenarios.

In the following section we describe a numerical method for transformation of probability densities  between the exterior- and interior-parameter spaces. In order to explore the distortion to the joint posterior distribution of EOS parameters incurred by being ignorant of singularities in the Jacobian, we will apply this numerical method as though the piecewise-polytropic EOS model defines an invertible mapping between the exterior- and interior-parameter spaces.  

\subsection{\textsc{MoRSE}: A post-processing code}
To infer EOS parameters from a posterior distribution of masses and radii we wrote a code named \textsc{MoRSE}.\footnote{Anagram of \textbf{M}ass and \textbf{R}adius to \textbf{E}quation \textbf{o}f \textbf{S}tate, available on \url{https://github.com/GRaaijmakers93/MORSE}.} \textsc{MoRSE} is a post-processing code in the sense that the input of the code is a joint posterior distribution -- the output of Bayesian estimation of exterior spacetime parameters. Here we will describe the workings of the code and discuss its numerical robustness. 

\subsubsection{Masses as integration variables}
\label{sec:rhotom}
Following OP09 we decrease computational expense by transforming onto the joint space of EOS parameters and \textit{masses} instead of onto the space of interior parameters (EOS parameters and central densities; the natural space to transform to): for the EOSs we consider there exists an invertible mapping from central densities to masses for a \textit{fixed} EOS $\boldsymbol{\theta}$. The vector of exterior parameters, $\bm{z}$, remains unchanged, as does the set $Z$ (the codomain of map $h$). However, the domain $Y$ and the vector of parameters, $\bm{y}$, are redefined such that $\bm{y}$ consists of $n$ EOS parameters and $s$ masses, and is thus interior-exterior mixed. To fully specify the EOS we still require that $s=n$, but now the Jacobian matrix contains fewer rows and columns: we only need to transform the radii into EOS parameters at a fixed triplet of masses. The elements in the Jacobian are filled by the partial derivatives of the function $R\left(\bm{\theta},M\right)$ with respect to EOS parameters, whilst the mass is kept fixed:
\begin{equation}
J(\bm{y}) = \left| \det \left(\frac{\partial R}{\partial P_i}\Biggr|_{M_j}\right) \right|.
\label{eqn:Jacobian partials}
\end{equation}
We can write the joint posterior distribution of the EOS parameters as
\begin{equation}
\label{PosteriorM}
\begin{aligned}
\mathcal{P}(\bm{\theta} ~|~ \mathcal{D}, \mathcal{M}, \mathcal{I}) \propto & \mathop{\int} \left[\prod_{i=1}^{s}\mathcal{P} (M_i, R_i ~|~ \mathcal{D}_{i}, \mathcal{M}, \mathcal{I})\right]J(\bm{y})d\bm{M}, \\
\end{aligned}
\end{equation}
where $R_{i}\coloneqq R(\boldsymbol{\theta},M_{i})$ and the upper limit\footnote{The lower limit is always a small non-astrophysical mass ($0.3$ M$_{\odot}$).} in each dimension of the integral is $M_{\textrm{max}}(\boldsymbol{\theta})$, the maximum mass permitted by the EOS $\boldsymbol{\theta}$ -- cf. Equation~(\ref{PosteriorRho}). In Appendix \ref{mvsrho} we show the numerical differences which manifest in the joint posterior distribution of the EOS parameters if one uses masses in lieu of central densities as the integration (marginalisation) variables.

The remainder of this section will focus on estimation of the EOS parameters $P_1$, $P_2$, and $P_3$, given the ansatz of an input joint posterior distribution of masses and radii being equal to a product of three (``constituent'') two-dimensional joint posterior distributions of mass and radius. The relevant joint posterior distribution of the EOS parameters is thus Equation~(\ref{PosteriorM}) with $s=3$.

\subsubsection{The EOS parameter space}\label{sec:EOS parameter space definition}
To numerically solve Equation~(\ref{PosteriorM}), we start by defining a three-dimensional grid of EOS parameters. The grid is calculated by taking all combinations of 50 logarithmically spaced points for each of the free parameters $P_1$, $P_2$, and $P_3$ from the following limits:
\begin{equation*}
\centering
\begin{gathered}
33.5 \leq \log_{10}{(P_1)} \leq 34.8, \\
34.5 \leq \log_{10}{(P_2)} \leq 36.0, \\
35.0 \leq \log_{10}{(P_3)} \leq 37.0.
\end{gathered}
\end{equation*}
The limits are chosen to be similar to those used in OP09, in order to allow for a comparison.\footnote{The precise range of parameters used in the OP09 study are not specified, so we adopt the range of parameters used in their plots. Some small differences between the results of their study and ours may be expected as a result.} Note that this range permits stars with maximum masses lower than 2 M$_\odot$: we have elected not to change this. Firstly, it enables us to compare our results properly with OP09 and hence verify that our code works as intended.
Secondly, excluding EOSs that do not permit a 2 M$_{\odot}$ star is more principled from the perspective of a Bayesian because these EOSs are still assigned a finite probability density and may, in principle, be strongly supported in future analyses conditioned on independent astronomical data sets. A physical reason to not neglect softer EOSs is the theoretical coexistence of distinct families of compact stars \citep[for recent papers on this topic see][]{Alford13, Drago2016a, Drago2016b, Alford17, Bhattacharyya17}: if the joint posterior distribution of EOS parameters is updated in the future, conditional on observations of distinct stars, and those stars do \textit{not} share an EOS with previously observed stars -- contrary to the modelling assumption of a \textit{shared} EOS -- then evidence for two families (subpopulations) may manifest as complex structure (e.g., multi-modality) in the updated joint posterior distribution. Moreover, it is generally the case that past analyses can \textit{in principle} be proven inaccurate to some degree (e.g., due to the existence of statistical bias) with appropriate model predictive checking and comparison \citep[see, e.g.,][]{Gelman_book}. 

We impose the condition that pressure must always be increasing with density to ensure that the EOS is microscopically stable \citep{Shapiro83}, meaning that $P_1 < P_2 < P_3$. Furthermore, we require that $0.5 \leq \Gamma \leq 6.5$ for all polytropic indices, based on the conclusions of \citet{Raithel16} but allowing for a slightly wider range. Finally, we test that each EOS is causal for a large range of central densities. The maximum density for which Equation~(\ref{causal}) still holds is taken as the maximum permitted central density given a EOS. 

We construct a grid of EOS parameter vectors because the general relativistic mapping is tractable to solve as $\boldsymbol{y}\mapsto\boldsymbol{z}$ -- i.e., in the forward local-to-global direction. In other words, we feign ignorance to mapping pathologies and fetch probability density from the space of exterior parameters, applying a differential volume transformation. It is important to note that this grid \textit{partially} specifies the EOS prior. When invoking the EP-paradigm, the prior is first specified on the space of exterior parameters and multiplied by a joint likelihood function. For the purpose of demonstration we have chosen an analytical \textit{posterior} distribution of the exterior parameters. The prior on the space of the interior parameters is then implicitly defined under the transformation in Equation~(\ref{PosteriorM}). Resultantly, a set of densities are associated with nodes of the grid after marginalisation over central densities; the grid resolution is therefore relevant for subsequent density interpolation. Moreover, the bounds of the grid control the bounds of the EOS prior and thus influence the iso-density contours of the transformed posterior distribution of EOS parameters. We therefore construct the grid such that it spans the largest possible set of EOS parameter vectors within the bounds set by any physical constraints (such as microscopic stability and causality).

\subsubsection{The Jacobian determinant}
\label{sec:jacobian}

\begin{figure*}
  \includegraphics[width=\textwidth]{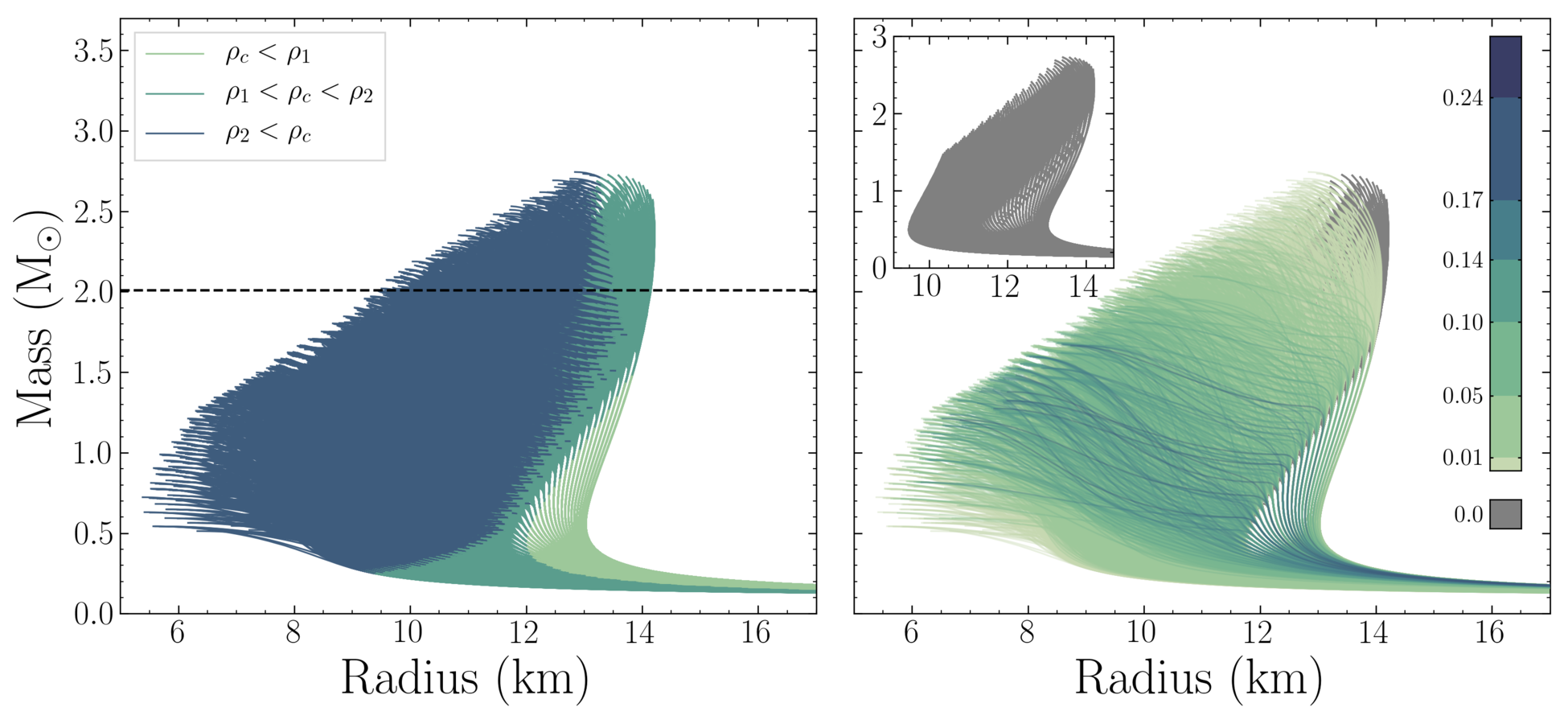}
  \includegraphics[width=1.1\columnwidth]{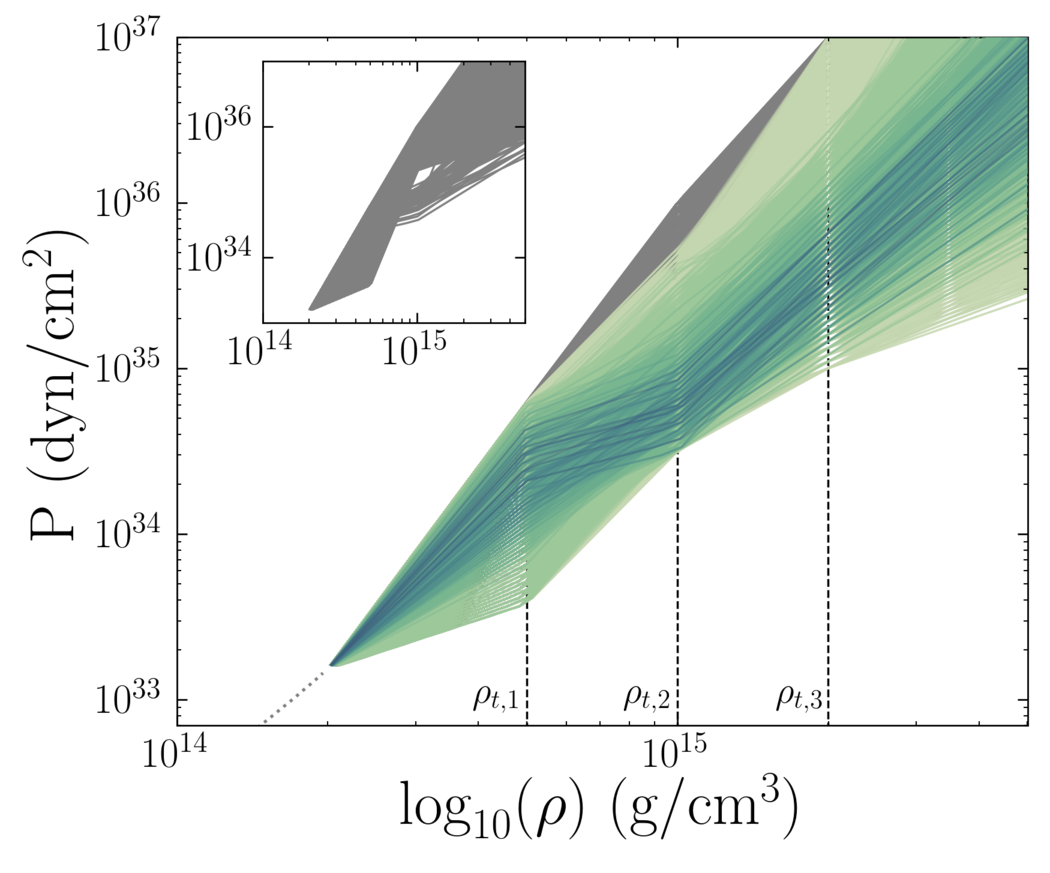}
  \caption{\textit{Left}: A set of sequences of \textit{stable} mass-radius solutions to the TOV equations -- one sequence (or curve) per EOS. Each curve is split into three coloured segments: each segment corresponds to exterior solutions which match to stable interior solutions (to the TOV equations) with a central density falling within the range of one of the three polytropes as given by Equation~(\ref{eqn:polytrope ranges}). For the stiffer EOSs, central densities in the range of the third polytrope contribute negligibly to the mass-radius curve. Note that, for the reasons outlined in the text, the range of EOS models considered permits masses below the current highest known pulsar mass of $2.01$ M$_\odot$ (black, dashed line). However, as is clear from the figure, the problems with polytropes not contributing to the overall $(M,R)$ curve are {\it more} acute for stiffer EOSs. \textit{Right}: The mass-radius curves are coloured according to the probability density of each associated EOS, under the ansatz that the input joint posterior distribution of masses and radii is jointly uniform. The darker curves are generated by EOSs with probability density -- see Equation~(\ref{eqn:Jacobian integral over mass}). Intermediate mass-radius curves that exhibit softening -- i.e., are quasi-horizontal at higher central densities -- are associated with larger integrals of the (modulus of the) Jacobian determinant over masses, than are softer and stiffer EOSs. Note that the stiffest EOSs have a probability density of zero (shown in grey) due to the Jacobian being everywhere singular. For clarity, the mass-radius curves whose generating EOSs have probability densities of zero are also shown in the inset panel. \textit{Bottom}: Similar to the right panel, but with the EOSs shown in the pressure-density space. The dotted line corresponds to the SLy EOS to which all polytropes are matched, and the dashed lines indicate the transition densities for the polytropes.}
  \label{fig:monotropes}
\end{figure*}

Given a library of EOSs which satisfy the aforementioned conditions (see Section \ref{sec:EOS parameter space definition}), together with their associated maximum central densities, a corresponding library of mass-radius curves are calculated by integrating the Tolman-Oppenheimer-Volkoff equations.\footnote{For most compact stars rotational corrections to the structure equations will be required. However our purpose here is to demonstrate an undesirable property of the piecewise-polytropic model, and for this purpose the non-rotating field equations suffice. These problems persist when the spacetime is rotationally deformed.} 
Note that since this coupled set of ordinary first-order differential equations is integrated radially outwards given a central density $\rho_c$, the star only contains matter at densities lower than $\rho_c$. In other words, the star's mass and radius are only dependent on the EOS at densities lower than $\rho_c$. In the piecewise-polytropic parametrisation, each parameter $\theta_i$ affects the EOS only at densities higher than the associated transition density $\rho_{t,i-1}$, as illustrated in Fig. \ref{fig:jacdeterminant}.

To compute a library of Jacobian determinants, the partial derivatives given by Equation~(\ref{eqn:Jacobian partials}) are first calculated for each EOS (in the EOS library), at 40 masses linearly spaced between 0.3 M$_{\odot}$ and 3.1 M$_{\odot}$. For one EOS and three stars the Jacobian is a three-by-three matrix, whose determinant is a function of the three masses. For each point $(P_1, P_2, P_3)$, a grid spanning a three dimensional subspace of masses is thus constructed, where each grid node is a point $(M_1,M_2,M_3)$; for each of these points on the grid a Jacobian determinant is calculated.\footnote{Note that each point in parameter space where two or three stars have equal mass values, the Jacobian is singular. Beyond the pathology deriving from the EOS parametrisation (Fig.~\ref{fig:jacdeterminant}) which for the purpose of demonstration is the main focus of this work, there evidently exist other pathologies in the interior-exterior mapping that one should be aware of. We direct the reader to R18 (sections 2.3.3 and 3, and appendix A) for further discussion on mapping pathologies.}

The choice of the piecewise-polytropic parametrisation (and thus a non-invertible mapping between the interior and exterior parameter spaces) results in a singular Jacobian over a fraction of the three-dimensional space of masses (or, equivalently, the three-dimensional space of central densities). To gain an understanding of the importance of this fraction, we examine the segments of the mass-radius curves generated by central densities in the domain of each polytrope in Fig. \ref{fig:monotropes}. When the EOS is stiff, central densities corresponding to the range affected by the third polytrope only generate a small segment of the mass-radius curve. As a result, a point in the three-dimensional mass space with all three mass values corresponding to central densities $\rho_c < \rho_{t,2}$ will always have a singular Jacobian. As long as the condition given by Equation~(\ref{eqn:non-singular condition}) is fulfilled, the Jacobian determinant will be finite. For the soft and intermediate EOSs the contribution of the third polytrope is significant compared to the total mass-radius curve, and thus the potential for distortion of the joint posterior distribution of EOS parameters is lower. Furthermore, we note that curves characterised by a long quasi-horizontal branch (softening) exhibit longer segments (in the astrophysical regime) which are dependent on the third polytrope ($P_3$). Thus, when conditioning on astronomical observations, the EOSs which generate such curves are not plagued by the existence of Jacobian singularities in astrophysical regions of the subspace of central densities or masses (to be marginalised over); these EOSs may thus be heavily favoured \textit{a posteriori} over, e.g., stiff EOSs. 

The effect of singularities in the Jacobian is to distort the joint posterior distribution of EOS parameters by superficially favouring certain EOSs over others. In order to explore this effect in more detail we perform the integral in Equation~(\ref{PosteriorM}), using a (bounded) joint uniform probability distribution over the entire parameter space for all three constituent posteriors, where each is denoted by $\mathcal{P} (M_i, R_i \,|\, \mathcal{D}_{i}, \mathcal{M}, \mathcal{I})=U(M_{i},R_{i})$. Doing so reduces Equation~(\ref{PosteriorM}) to a three-dimensional integral over the subspace of masses for a fixed EOS $\boldsymbol{\theta}$:
\begin{equation}
\mathcal{P}(\bm{\theta} ~|~ \mathcal{D}, \mathcal{M}, \mathcal{I}) \propto \mathop{\int} U(\boldsymbol{z})J(\bm{y})~dM_1 dM_2 dM_3.
\label{eqn:Jacobian integral over mass}
\end{equation}
The joint posterior distribution of the EOS parameters that follows from this calculation allows us to examine the degree to which the relative posterior weightings are distorted by the singularities. 
In the right panel of Fig. \ref{fig:monotropes} we plot the mass-radius curves, colour-coded by the probability density of the EOS parameters that generate the curve. The figure illustrates that the (intermediate) mass-radius curves generated by EOSs which exhibit softening in the domain of the second polytrope (that is, curves with an extended quasi-horizontal segment) are favoured over softer or stiffer curves.\footnote{A complicating factor is that the subset of EOSs which exhibit second polytrope softening (see the dark, coloured EOSs in Fig. \ref{fig:monotropes}) generate exterior spacetime solutions which are highly sensitive to variations in $\bm{\theta}$ near the turn-over points in the mass-radius plane. In particular, partial derivatives of the radius with respect to the pressures can be relatively high, and thus the Jacobian determinant can be large. To prevent this effect, one could adjust the range of EOS parameters in order to discard such mass-radius curves: e.g., one could increase the separation between the upper limit of $P_1$ and the lower limit of $P_2$. However, doing so would omit EOSs that exhibit a constant pressure phase transition at these densities and thus the full set of microscopically stable candidate EOSs would not considered.}

\begin{figure*}
    \includegraphics[width=\textwidth]{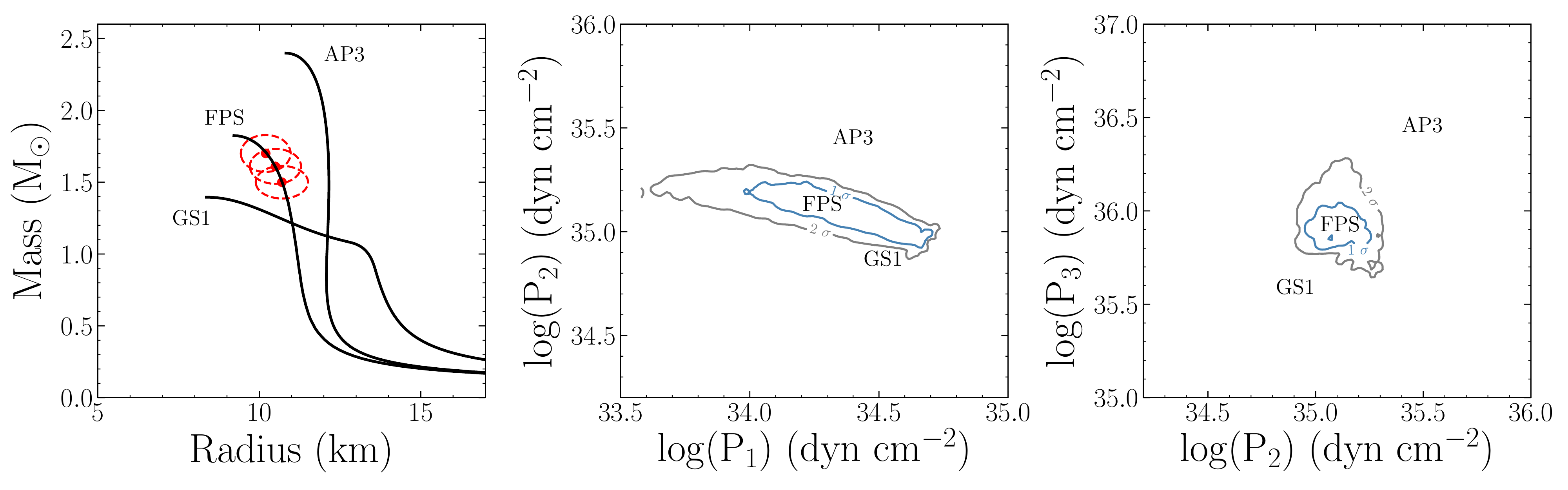}
    \caption{In the left panel a two-dimensional projection of the joint posterior of masses and radii used in OP09 to infer the EOS parameters, centered on the FPS EOS, which we replicate here (compare to Figure 3 of their paper). The dashed red ellipses represent the $1\sigma$ credible regions of the constituent posteriors of mass and radius. For comparison, two other EOSs, AP3 and GS1, are also displayed. The panels on the right show the posterior distribution of the EOS parameters inferred with \textsc{MoRSE}, which should be compared to Figure 5 of OP09. All distributions are marginalised over the parameter not shown. Our distributions are similar, although not identical, to those of OP09: we believe that the differences are most likely due to differences in the applied computational methods (see the discussion in the main text). However, the calculations are consistent in terms of distinguishing the thermodynamical manifestations of three microphysical models (FPS, AP3, and GS1).}
\label{CompareOzel} 
\end{figure*}

Our library of Jacobian determinants spans a three-dimensional grid of EOS parameters $P_1$, $P_2$, and $P_3$. Each node on this grid contains a nested three-dimensional grid of masses, and each node of this nested grid has an associated Jacobian determinant. The joint posterior distribution of the EOS parameters given by Equation~(\ref{PosteriorM}) is calculated separately for each grid-point $(P_1,P_2,P_3)$. First we construct a linear interpolator of the Jacobian determinant on the nested grid of masses -- i.e., an approximating function that takes three mass coordinates and outputs a determinant. For each grid-point $(P_1,P_2,P_3)$ we marginalise over the mass subspace to obtain a posterior distribution of the EOS parameters; the integrand is the (approximate) Jacobian determinant multiplied by a local joint posterior density of mass and radius -- see Equation~(\ref{PosteriorM}).

\subsubsection{Testing the code}
To test \textsc{MoRSE} we first attempted to reproduce the calculations of OP09. We use a similar parametrisation to OP09 and the same method of EOS parameter estimation. OP09 define for each star, a joint posterior distribution of its mass and radius, with a mode which is a stable solution to the TOV equations given the FPS EOS \citep{Lorenz93}. Analytically, the joint posterior distribution of masses and radii is written as the product of three two-dimensional Gaussians:
\begin{equation}
\label{GaussiansOzel}
\mathcal{P}(\boldsymbol{z}~|~\mathcal{D}, \mathcal{M}, \mathcal{I}) = \prod_{i=1}^3 \exp\left[-\frac{(M_i - \widehat{M}_{i})^2}{2 \sigma_{M,i}^2} -  \frac{(R_i - \widehat{R}_{i})^2}{2 \sigma_{R,i}^2}   \right]\,.
\end{equation}
The modes are chosen to be: $\widehat{M}_{1} = 1.5$ M$_{\odot}$, $\widehat{R}_{1} = 10.7$ km; $\widehat{M}_{2} = 1.61$ M$_{\odot}$, $\widehat{R}_{2} = 10.5$ km; $\widehat{M}_{3} = 1.7$ M$_{\odot}$, $\widehat{R}_{3} = 10.2$ km. The distribution $\mathcal{P}(\boldsymbol{z}\,|\,\mathcal{D}, \mathcal{M}, \mathcal{I})$ is thus a six-dimensional function of the exterior parameters $(M_1, M_2, M_3, R_1, R_2, R_3)$. The standard deviations of each Gaussian are set to $5 \%$ of the modal parameter values.  In Fig. \ref{CompareOzel} we project the constituent joint posterior distributions onto the same plane, together with the mass-radius curves generated by the FPS EOS and two other EOSs \citep{GS1, AP3}. These three EOSs together span a wide range of thermodynamic conditions, and span a wide range of masses and radii.  

The six-dimensional joint posterior distribution of the masses and radii is transformed using Equation~(\ref{PosteriorM}) to obtain a joint posterior distribution of the EOS parameters. Following OP09 we plot this distribution in a two-dimensional plane, where the parameter not shown is marginalised over. In Fig. \ref{CompareOzel} we show the results obtained with the \textsc{MoRSE} code. OP09 concluded that a $5 \%$ uncertainty in the Gaussian mass-radius posterior distributions is enough to distinguish on a $2\sigma$ credible region between three distinct EOSs with different underlying microphysics. The joint posterior distributions produced with \textsc{MoRSE} show a similar contour structure to those of OP09, although there are discernible differences. We believe these differences are most likely attributable to factors such as different choices for the grid resolution in EOS parameter space, the methods of interpolation, and the integration routine. 

For the scenario considered here, one would expect the joint posterior distribution of the EOS parameters to be negligibly distorted due to the existence of singularities in the Jacobian. Most of the probability mass of the three Gaussian mass-radius posterior distributions is localised in a region where exterior solutions are generated by central densities in the range of the third polytrope. The Jacobian determinant is zero only where the joint posterior distribution of masses and radii exhibits negligible support. We expect, however, that if the modes of the Gaussian mass-radius posteriors were exterior solutions generated by an FPS EOS with lower central densities, or were exterior solutions generated by a stiffer EOS, the incurred distortion would be more significant. We now go on to explore this forecast in more detail.

\subsection{Configurations of the input joint posterior distribution of masses and radii}
\label{subsec:scenarios}
In order to explore the distortion of the joint posterior distribution of the EOS parameters, we perform transform a several different joint posterior distributions of masses and radii, spanning a large range of the possible (six-dimensional) parameter space. Our expectation from R18 is that if one is ignorant of the existence of singularities in the Jacobian, one can in principle appreciably distort the posterior distribution of the EOS parameters. The constituent joint posteriors (one per star) we consider are bivariate Gaussian distributions with varying standard deviations and modes -- see Equation~(\ref{GaussiansOzel}). The modes of these constituent joint posterior distributions are always defined to be stable exterior spacetime solutions generated by some ``underlying'' EOS.

\begin{figure}
    \includegraphics[width=\columnwidth]{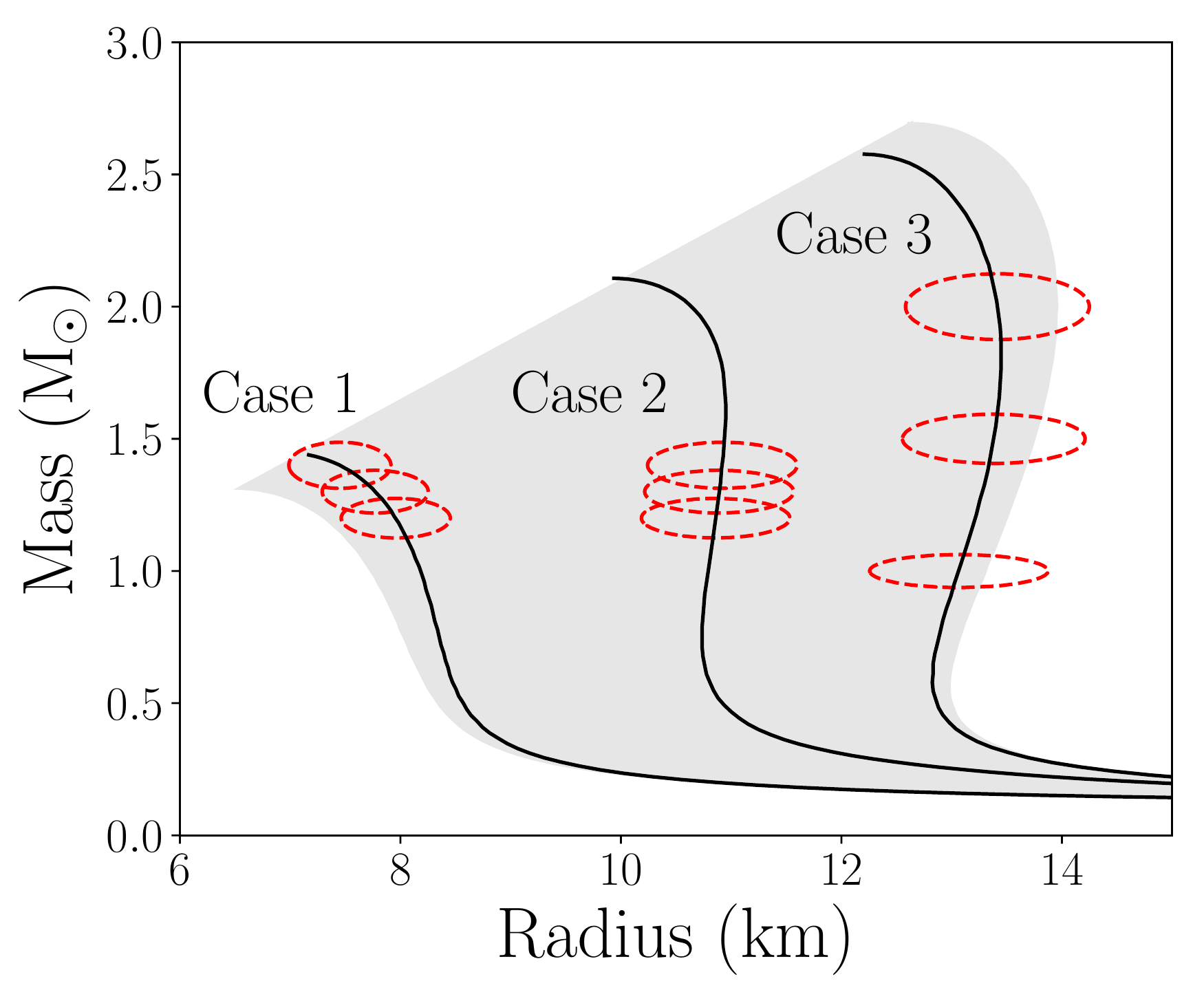}
    \caption{The shaded area is a continuous approximation to the region spanned by our library of mass-radius curves, i.e., the shaded region represents (a two-dimensional projection of) the image of the domain $Y$ under the map $h$ (see Section \ref{sec:parameterest}). The codomain $Z$ (of map $h$), which a joint probability distribution of exterior parameters continuously spans, is a \textit{superset} of the shaded region. The black lines are three representative (or ``underlying'') mass-radius curves on which the modes of the constituent mass-radius posterior distributions are centred. The red dashed contours show the 2$\sigma$-levels of these constituent distributions for three possible configurations of the input joint posterior distribution of masses and radii. Cases 1 and 2 (see the main text) are applied to all three representative EOSs, while Case 3 is only applied to the intermediate and stiff EOSs.}
    \label{fig1}
\end{figure}

In order to reduce the required computational time, we limit the number of ``underlying'' EOSs  considered to three. To still examine a broad range of mass-radius curves, we choose three representative EOSs that differ significantly from each other: the first is a \textit{soft} EOS with a maximum mass of $\sim1.4$ M$_{\odot}$; the second is an \textit{intermediate} EOS with a maximum mass of $\sim2.1$ M$_{\odot}$; and the third is a \textit{stiff} EOS with a maximum mass of $\sim2.6$ M$_{\odot}$ (Fig. \ref{fig1}). One can argue that given the discovery of a $2.01$ M$_{\odot}$ star \citep{Antoniadis13}, the first (soft) EOS could be discarded immediately. We have already explained why we do not discard EOSs which do not meet the $\sim2$ M$_{\odot}$ in Section~\ref{sec:EOS parameter space definition}; such soft EOSs also serve to make clear distributional distortion incurred in the stiff regime of the EOS parameter space.

\begin{figure}
  \includegraphics[width=\columnwidth]{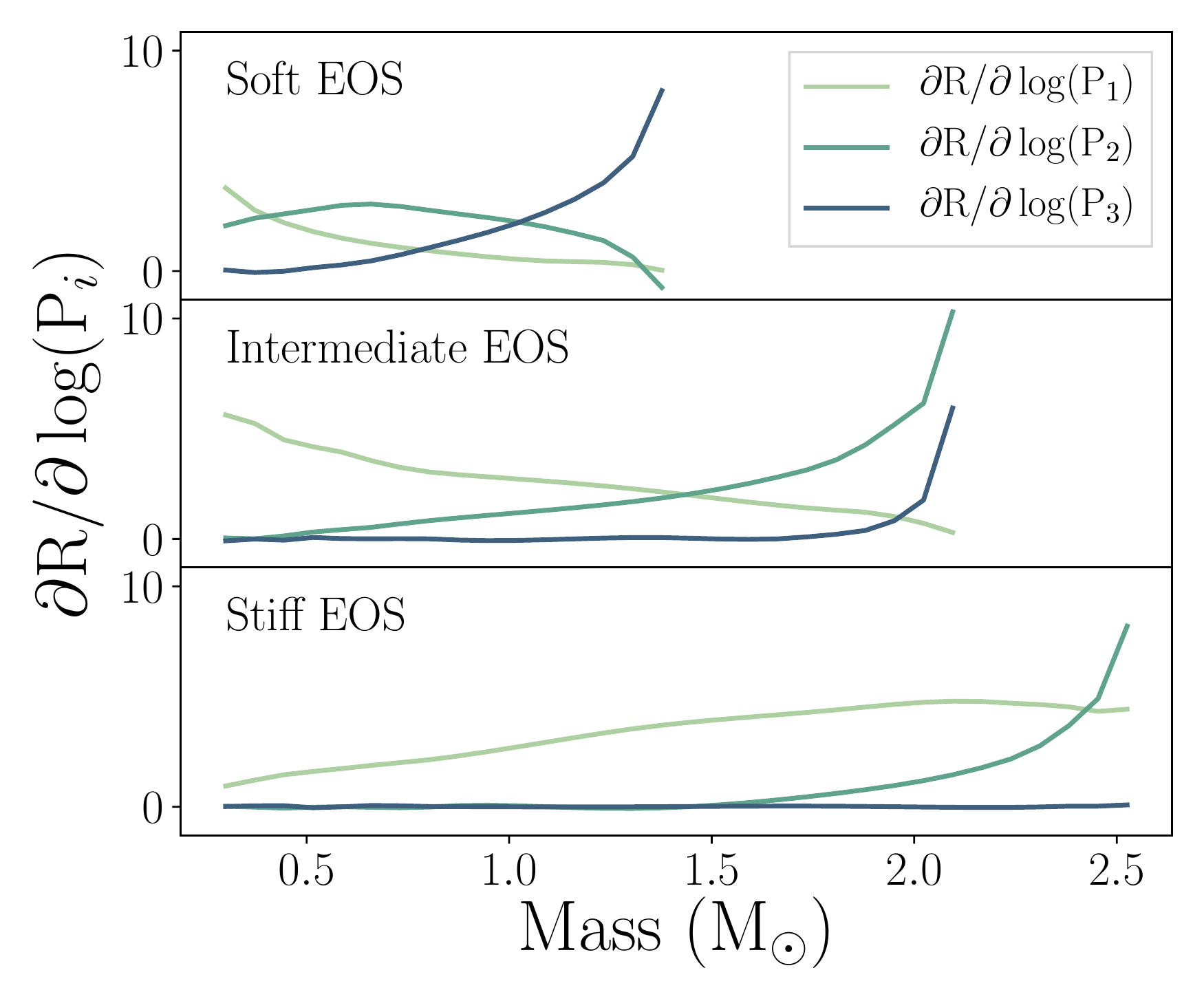}
  \caption{The partial derivatives $\partial R / \partial \log_{10}{P_i}$, for the three representative EOSs, as a function of mass. For both the intermediate and stiff EOSs, the partial derivative of radius with respect to $P_3$ is (effectively) zero for a large range of astrophysical masses.}
  \label{fig:partialderiv}
\end{figure}

For each underlying EOS, we compute the partial derivatives that enter in the Jacobian matrix for all stable values of the mass; we then evaluate whether or not the Jacobians is anywhere singular in the astrophysical mass regime (Fig. \ref{fig:partialderiv}). Let us first consider the soft EOS: the Jacobian is non-singular in the astrophysical regime of the three-dimensional mass subspace (to be marginalised over), because the partial derivative of the radius with respect to $P_3$ is only zero for masses below ${\sim} 0.5$ M$_{\odot}$. However, for the intermediate and the stiff EOSs, the partial derivatives of radius with respect to $P_3$ are zero for masses below ${\sim} 1.7$ M$_{\odot}$ and ${\sim} 2.5$ M$_{\odot}$ respectively -- that is, in the astrophysical regime. If, for each star, there is no posterior probability mass at \textit{gravitational} masses greater than ${\sim} 2.5$ M$_{\odot}$, then: the stiff EOS will not be assigned a finite posterior probability -- despite being potentially permitting strongly favoured exterior solutions -- because of singularities in the Jacobian.\footnote{We note that due to small numerical inaccuracies, Jacobian determinants calculated with \textsc{MoRSE} are never exactly zero. This is because each mass-radius curve is calculated for different central densities, even when, e.g., the parameters $P_1$ and $P_2$ are shared between curves. As a result, the function $R\left(\bm{\theta},M\right)$ is never truly constant and thus the partial derivatives never truly zero but of $\mathcal{O}(10^{-14})$.} In summary, joint posterior distributions of masses and radii will be distorted under transformation, provided that exterior spacetime solutions are supported in the region spanned by the (astrophysical) solutions for intermediate and stiff EOSs.

The mode of each constituent mass-radius posterior distribution is an exterior spacetime solution permitted by \textit{one} of the three underlying EOSs, and that underlying EOS is shared by the three modes. For the soft EOS, two different scenarios are considered; for the intermediate and stiff EOSs, on the other hand, three scenarios are considered. Two of these scenarios are characterised by the modes of the constituent posteriors being clustered together, either high on a mass-radius curve (Case 1) or in the middle of a curve (Case 2). In other words, similar exterior spacetime solutions are supported \textit{a posteriori} for all three stars. The third scenario is characterised by modes of the constituent posteriors being spread out over a mass-radius curve (Case 3). In other words, appreciably different exterior spacetime solutions are supported \textit{a posteriori} for all three stars. Specifically, for Case 3, we impose that the modal masses are $1.0$, $1.5$ and $2.0$ M$_{\odot}$, and that the modal mass-radius pairs each coincide with an exterior solution generated by the underlying intermediate and stiff EOSs. The three underlying EOSs and the Cases are illustrated in Fig. \ref{fig1}. Note that the modal exterior spacetime solutions are not \textit{chosen} in a real analysis, but are inferred conditioned on: a model (e.g., a numerical, statistical model for generation of an astronomical data set); an astronomical data set; and any prior information (independent data sets). 

For each of these Cases, the standard deviation of the bivariate Gaussian distribution is varied. Three variations are considered where the constituent joint posterior distributions have equal standard deviations in mass and radius of $10$, $5$ or $2.5 \%$ of the modal values. These variations, applied to the Cases described above, generate a set of $24$ \textit{scenarios} for which the joint posterior distribution of the EOS parameters is computed.
 
\subsection{Measures for assessing distributional distortion}
\label{subsec:quality}

An exact metric for quantification of induced distributional distortion does not exist: an ``exact'' (non-distorted) joint posterior distribution of the EOS parameters for each scenario is not calculable because the mapping between interior and exterior parameter spaces is non-invertible (and thus joint prior distributions cannot be transformed between parameter spaces; see Section \ref{sec:parameterest} for more detailed reasoning). It follows that the joint posterior distributions calculated by \textsc{MoRSE} cannot be compared to some ``exact'' distribution of the EOS parameters in order to fully quantify the distortion. In the following paragraphs we discuss two possible measures that allow us to infer that distortion has been incurred due to our (feigned) ignorance of the existence of Jacobian singularities.

Two crude scalar summary statistics for (unimodal) distributions on a parameter space are the MAP (maximum \textit{a posteriori}) parameter vector, and the expected (mean) parameter vector (an integral over the distribution). These scalar summaries alone are typically noninformative because many non-negligible distributional moments (or statistics) are neglected. We thus use the MAP and expected parameter vectors in combination with other measures to detect any distributional distortions.

An intuitive measure -- used for example by OP09 -- is to determine if the parameters of an underlying EOS are encompassed by the $1\sigma$  or $2\sigma$ credible regions of the joint posterior distribution of EOS parameters, i.e., the highest posterior density credible region. As we have chosen MAP mass-radius pairs which are exterior spacetime solutions generated given an underlying EOS, we will consider this underlying EOS as a reasonable approximation to the true EOS assumedly shared by the observed stars.\footnote{Joint posterior distributions of mass and radius inferred with current observational techniques will not in general support MAP mass-radius pairs which are permitted by some ``exact'' EOS (assuming the EOS model is exact and the observed stars are static general relativistic stars); nor will the distribution be necessarily well-approximated by a multivariate Gaussian distribution. Reasons for this are because of the existence of statistical biasing due to other inaccurate (nuisance) facets of forward models of data generation, and existence of complex posterior structure such as multi-modality and degeneracy which require many distributional moments to accurately represent. For a marginally more realistic approach, the modes of the joint posterior distributions could be scattered around an underlying EOS.} The $1\sigma$  and $2\sigma$ credible regions are defined by the isodensity contours that encompass $68.3$ and $95.5 \%$ of the total probability mass, respectively. Numerically, these contours are calculated by taking all sets of EOS parameters that account for $68.3$ and $95.5 \%$ of the total probability mass. This is effectively a one-dimensional optimisation problem: (i) we start with the maximum probability density in the distribution and iteratively lower this probability density; (ii) at each iteration we check what percentage of the total probability density is contained and stop when $68.3 \%$  (or $95.5 \%$) is reached. 

The above measures for identifying distortion are defined on the EOS parameter space. For astrophysicists, casting this information on the space of exterior parameters may prove useful. One way to achieve this is to calculate an (equally weighted\footnote{One could also weight according to the posterior distribution, but here we simply use the discrete subset of EOSs that are supported within the $1\sigma$ credible region.}) average offset in radius (at a fixed mass) between (i) an underlying mass-radius curve, and (ii) the mass-radius curve generated by an EOS parameter vector with finite posterior support within the $1\sigma$ credible region (an example of this measure can be seen in the right panels of Fig. \ref{ExamplesScenario}). Numerically, we take $250$ uniformly-spaced masses and find, for each mass, the corresponding radii for the mass-radius curves generated by EOS parameter vectors within the $1\sigma$ credible region. For these sets of radii we then calculate the absolute percentage offset from the radius of the underlying EOS, and average over these percentages. This method gives an indication of the typical deviation from the underlying mass-radius curve exhibited by EOS parameters bounded by the $1\sigma$ credible region. Note that as we take the average of the absolute percentage offset, this measure is never precisely zero, since the mass-radius curves within the $1\sigma$ credible region are distributed around the underlying curve.

\section{Results}
In Section~\ref{sec:demonstration of distortion} we perform a selection of calculations to demonstrate the problem of posterior distributional distortion, which can be incurred if a pathological mapping between interior and exterior parameter spaces is naively used. In Section~\ref{sec:extramass} we include a radio pulsar mass constraint to demonstrate that this information does not mitigate the problem.

\label{sec:results1}

\subsection{Demonstration of distortion}\label{sec:demonstration of distortion}
First we test whether an underlying EOS parameter vector lies within the $1\sigma$ or $2\sigma$ credible region of the joint posterior distribution, in scenarios with $\sigma_M$ and $\sigma_R$ set to $5 \%$ of modal values for each star. We display these calculations in Table \ref{tab1}.

We find that there are scenarios for which distortion is clearly incurred: in particular, if the modal mass-radius pairs are located in the middle segment of a mass-radius curve generated by an intermediate or stiff underlying EOS (scenarios under Case 2). This is because the underlying EOS is not recovered: there is relatively small support \textit{a posteriori} for the underlying EOS. Closer inspection of the joint posterior distributions of the EOS parameters reveals that for all of these scenarios, the values of $P_1$ and $P_3$ are within the $2\sigma$ credible region, while $P_2$ is not. The fact that $P_1$ is more tightly constrained by constituent posteriors with modes at lower masses can be intuited, since $P_1$ principally controls mass-radius curves at lower masses (lower central densities). The pressure $P_3$ is not constrained at all by posteriors with lower mass modes, so the underlying value of $P_3$ lies within the  $2\sigma$ credible region. The reason that $P_2$ is not within the  $2\sigma$ credible region is because the mass-radius curves with a  long quasi-horizontal branch, caused by soft value for $P_2$ (shown in Fig. \ref{fig:monotropes}), are strongly supported \textit{a posteriori}; the Jacobian determinant for such EOS parameter vectors is non-zero over a large range of astrophysical masses, as explained in Section \ref{sec:jacobian}. Similar reasoning can be used to explain why the underlying soft EOS parameters are not within the $1\sigma$ credible region for Case 2, but as the effect of Jacobian singularities is smaller for softer EOSs, the parameters do reach within the $2\sigma$ credible region.

The Case 3 scenario for the stiff EOS demonstrates that even when the mass-radius pairs are spread over a large segment of the mass-radius curve, the underlying EOS parameter vector is not recovered within the $1\sigma$ credible region. The reason for this is the Jacobian determinant being zero over a significant fraction of the the mass-radius curve, in combination with the fact that the configuration of constituent mass-radius posteriors supporting EOSs which are softer in the domain of the second polytrope. The underlying EOS parameter vector is only very weakly supported \textit{a posteriori}, with the lying within the $2\sigma$ credible region.

\begin{table}
\centering
\caption[]{In the table below we denote by a tick the scenarios for which the underlying EOS parameter vector is within the $1\sigma$ credible region (or, in brackets, the $2\sigma$ credible region) of the joint posterior distribution of EOS parameters. These scenarios are defined by the constituent mass-radius posteriors having standard deviations of $\sigma_M$ and $\sigma_R$ set to $5 \%$ of the modal values.}
\label{tab1}
\begin{tabular}{c|ccc}
\toprule
\textbf{Scenario} & \textbf{Soft EOS} & \textbf{Intermediate EOS} & \textbf{Stiff EOS} \\ \bottomrule
\toprule
Case 1   & \checkmark (\checkmark)   & \checkmark (\checkmark) & \checkmark (\checkmark)  \\
Case 2  & \xmark (\checkmark)  & \xmark (\xmark)  & \xmark (\xmark) \\
Case 3 & $-$  & \checkmark (\checkmark)   & \xmark (\checkmark)            
\end{tabular}
\end{table}

\begin{figure*}
    \includegraphics[width=\columnwidth]{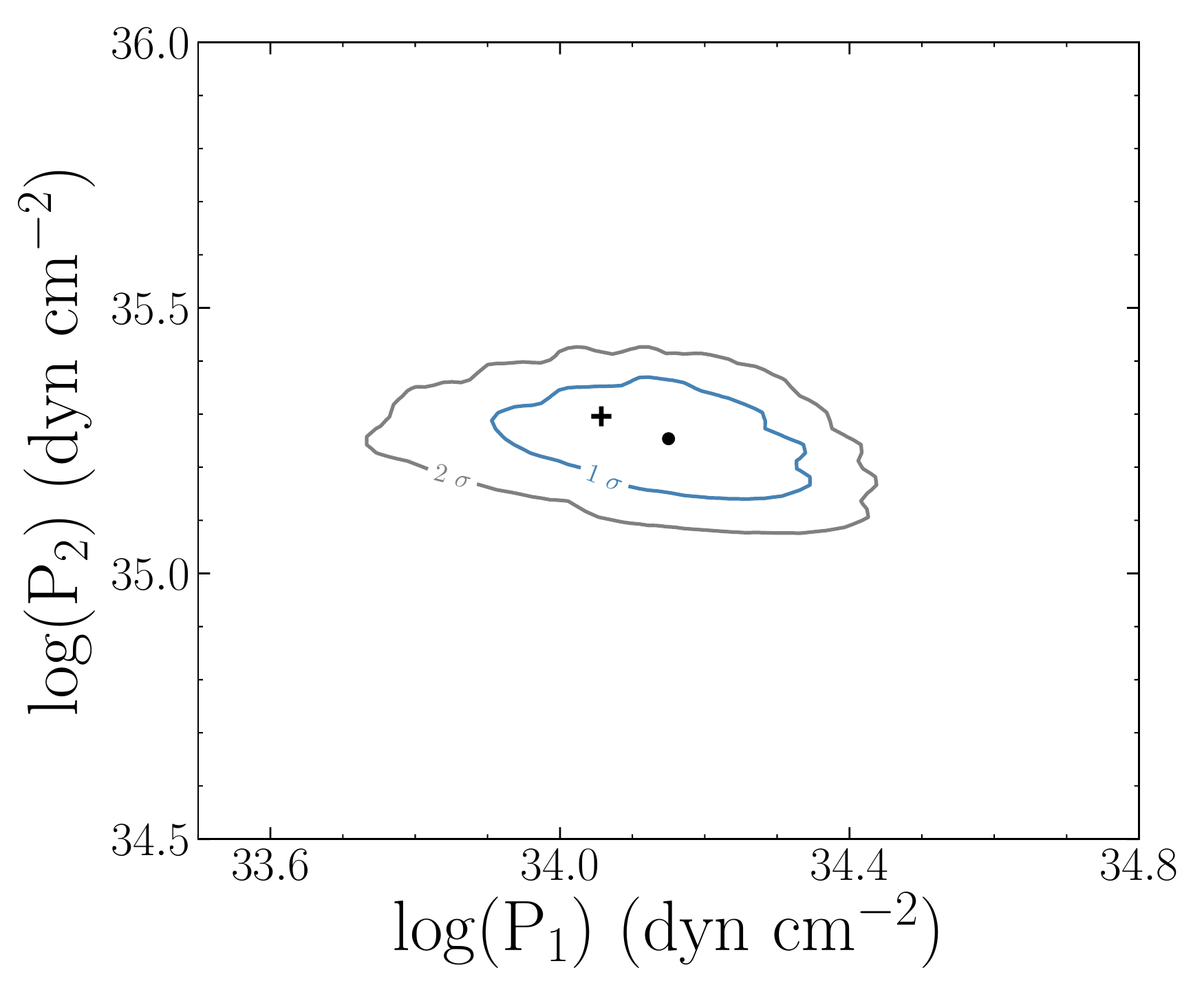}
    \includegraphics[width=\columnwidth]{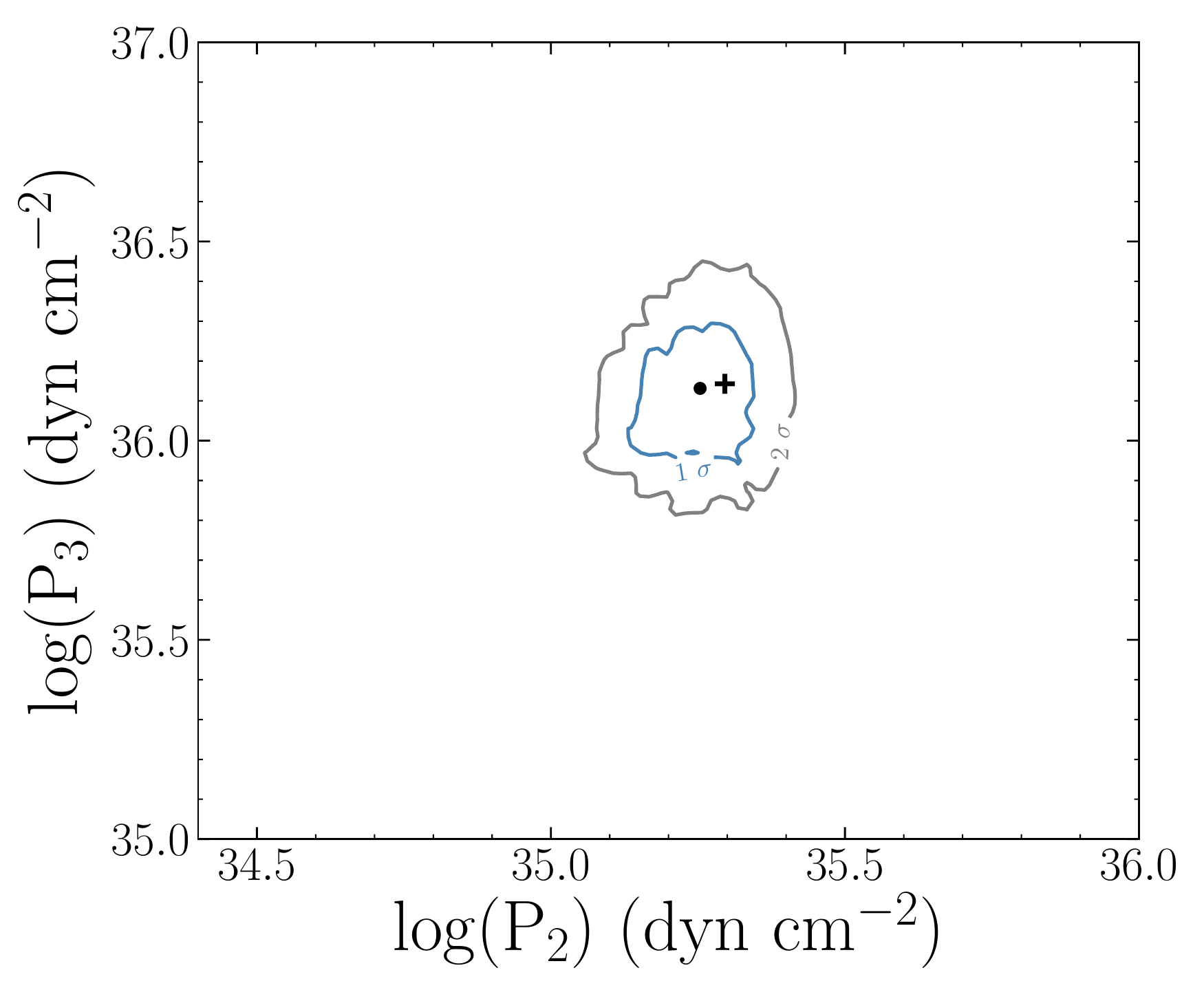}
    \includegraphics[width=\columnwidth]{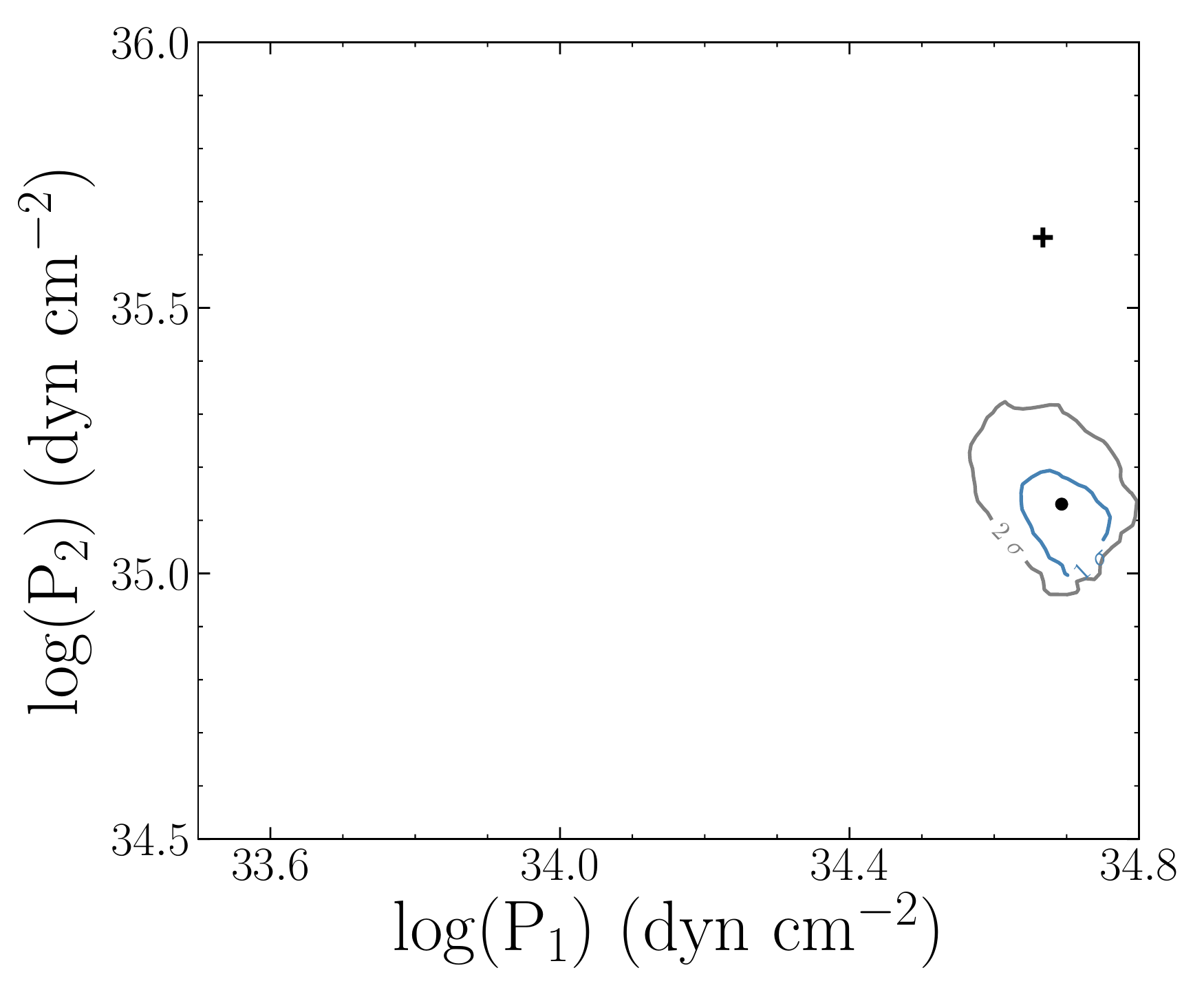}
    \includegraphics[width=\columnwidth]{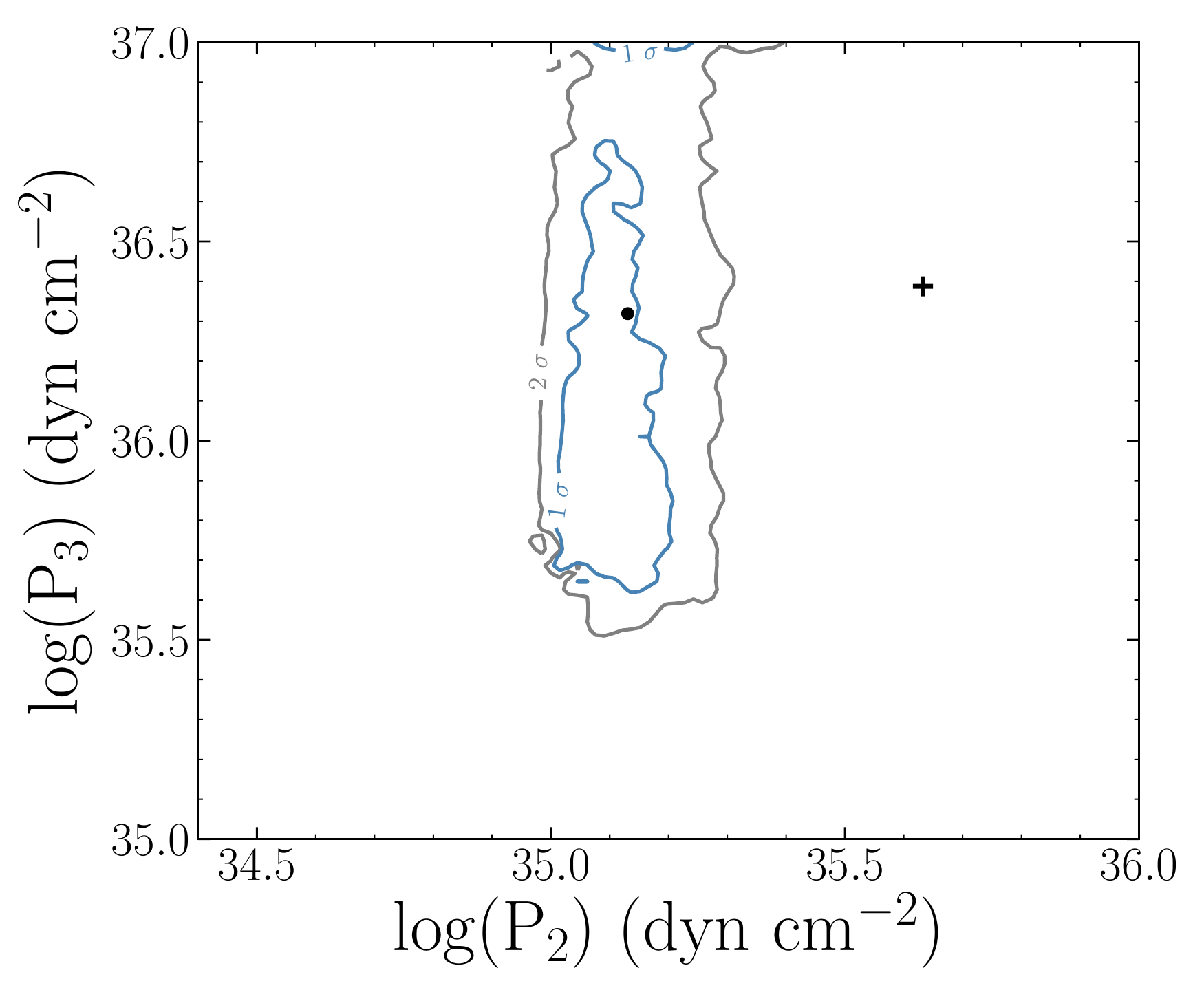}
  \caption{In this figure we show the posterior distribution of the EOS parameters for our two example scenarios, where the upper and lower panels are the resulting distributions of inferring from Case $3$ on the intermediate EOS and Case $2$ on the stiff EOS respectively. The blue and gray contours show the $1\sigma$ and $2\sigma$ credible region, the plus sign indicates the underlying EOS parameters and the dot shows the average EOS parameters. The grey area shows the full set of EOS parameters considered. We see that in the first scenario all EOS parameters are relatively well-constrained, with a small offset between the average and underlying EOS parameters. For the second scenario the value of $P_2$ shows extreme softening, leading to an unconstrained value of $P_3$.}
  \label{ProbEOS}
\end{figure*}

To further characterise distributional distortion, we proceed to analyse two example scenarios in greater detail. The first scenario is Case $3$, with modal mass-radius pairs permitted by the intermediate EOS. The second scenario is Case $2$ with modal mass-radius pairs permitted by the stiff EOS. In both scenarios the constituent mass-radius posteriors are all assigned standard deviations of $5 \%$ of their modal parameter values. We show the joint posterior distributions of the EOS parameters generated by \textsc{MoRSE} in Fig. \ref{ProbEOS}: the upper and lower panels respectively corresponding to the first and second scenarios. This figure illustrates that: (i) in the first scenario, the posterior-averaged EOS parameter vector is comparable to the underlying EOS parameter vector; and (ii) in the second scenario, the posterior-averaged EOS parameter vector is appreciably different to the underlying EOS parameter vector because the average pressure $P_2$ is much lower than the $P_2$ pressure of the underlying EOS. 

\begin{figure*}
    \includegraphics[width=.9\columnwidth]{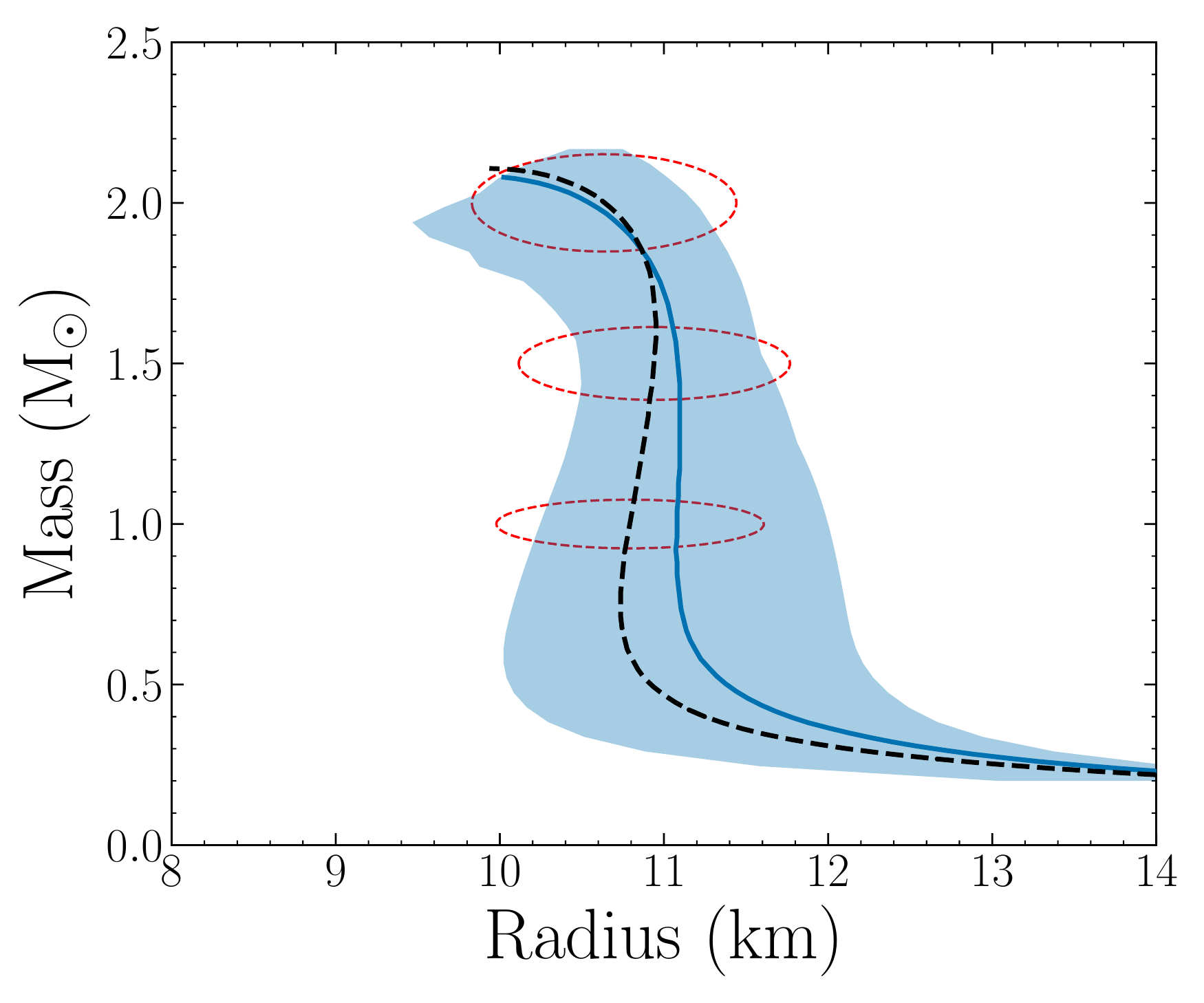}
    \hspace{1cm}
    \includegraphics[width=.9\columnwidth]{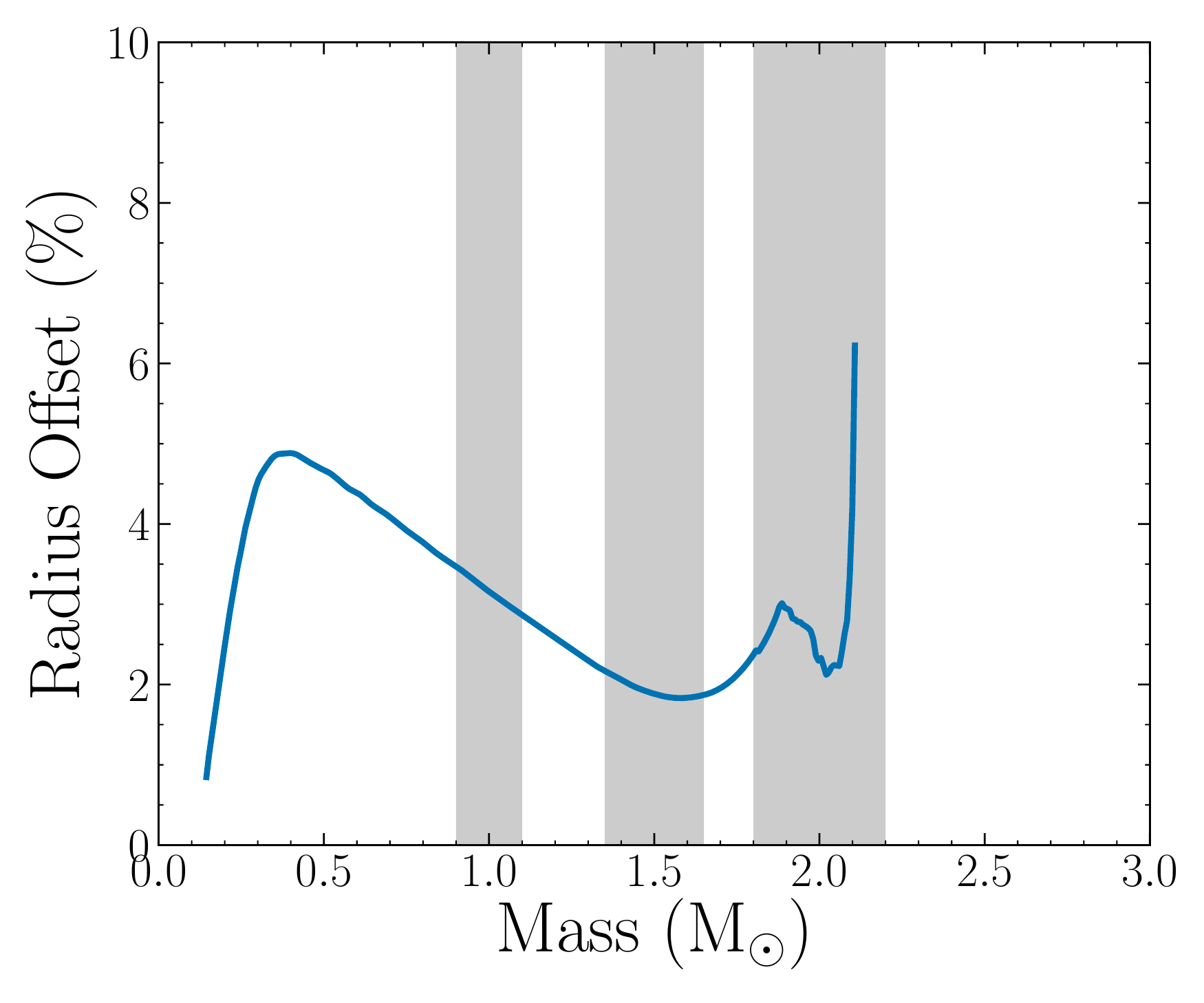} 
    \includegraphics[width=.9\columnwidth]{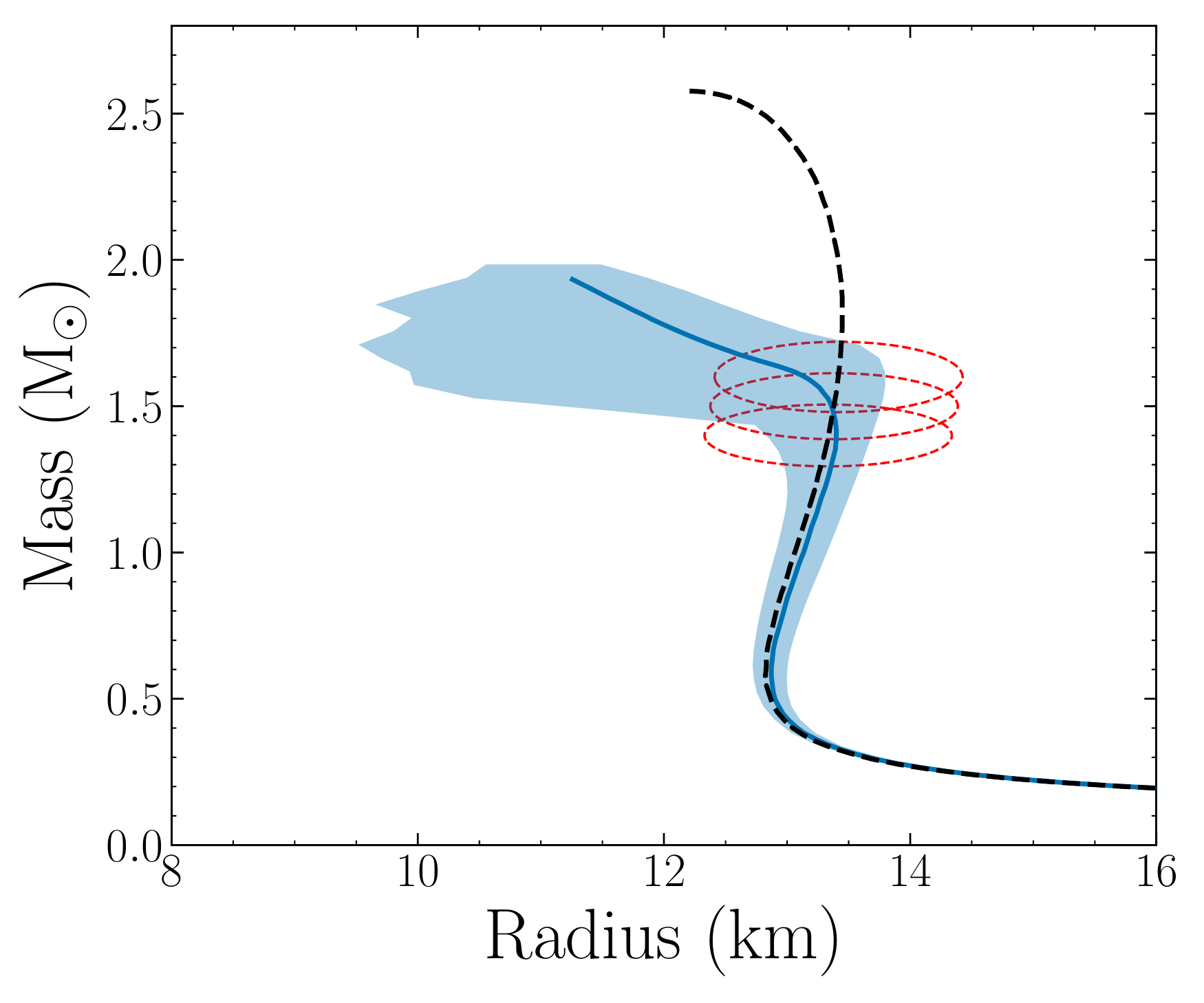}
    \hspace{1cm}
    \includegraphics[width=.9\columnwidth]{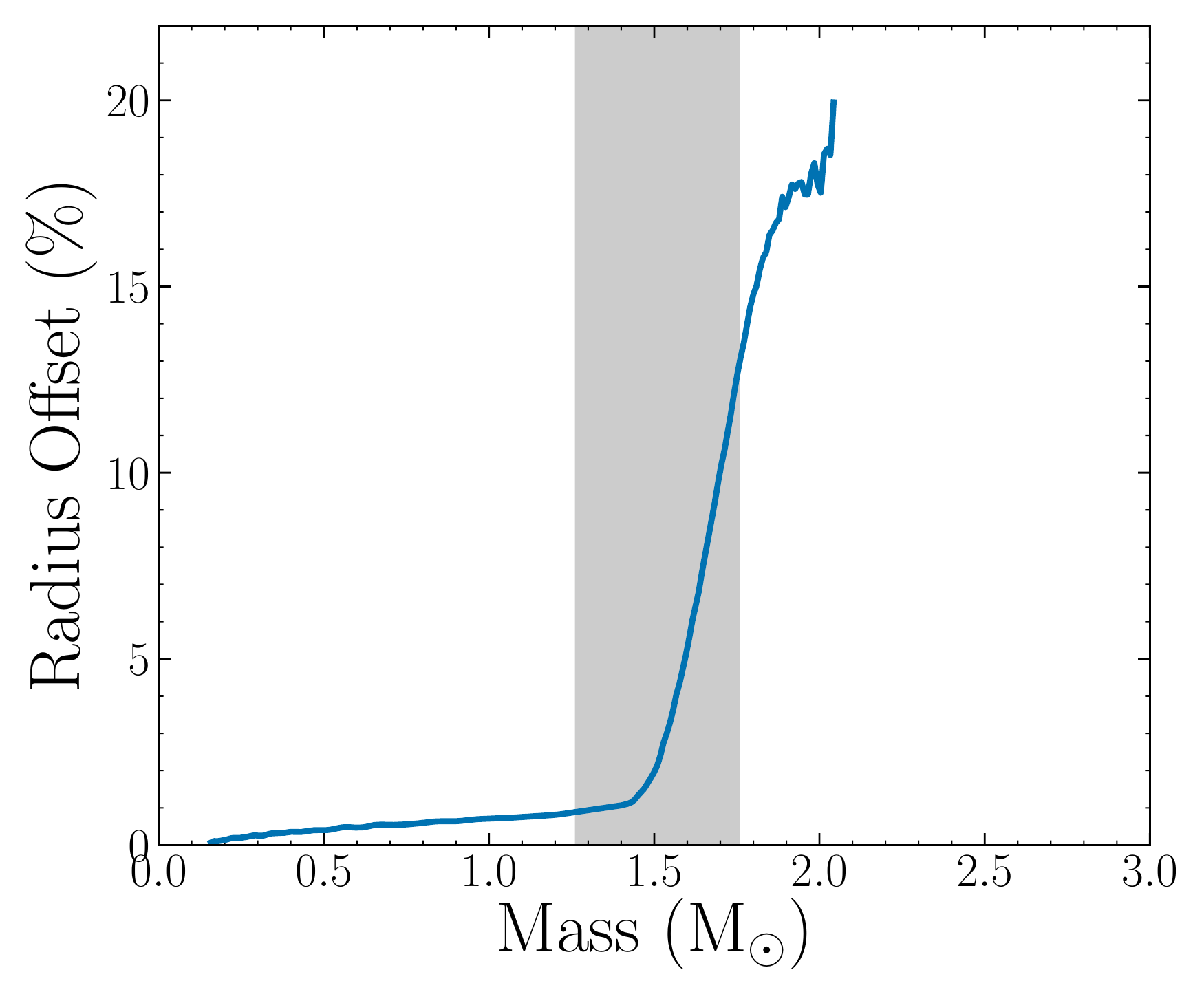} 
  \caption{The upper- and lower-panels correspond to the scenarios Case $3$ on the intermediate EOS and Case $2$ on the stiff EOS respectively. The solid blue curve is the posterior-averaged mass-radius curve. In the left panels the shaded areas indicate the region spanned by mass-radius curves generated by EOSs within the $1\sigma$ credible region. The black dashed curves show the underlying mass-radius curves (on which the modal mass-radius pairs are chosen to lie). The red dashed ellipses are projections of the $1\sigma$ contours of the \textit{constituent} mass-radius posteriors. In the upper-panel the underlying mass-radius curve is recovered within the shaded region and is everywhere comparable to the posterior-averaged mass-radius curve. Conversely, in the lower-panel, the underlying mass-radius curve is not everywhere comparable to the posterior-averaged mass-radius curve. The favoured EOSs are softened in the domain of the second polytrope relative to the underlying EOS, as seen in Fig. \ref{ProbEOS}. Similar conclusions can be reached by examining the right panels: the offset in radius is plotted for both scenarios (as explained in Section \ref{subsec:quality}). Note that in the upper-right panel, the behaviour of the solid curve beyond masses of $\sim2$ M$_{\odot}$ (local minimum and sharp rise) manifests because an increasingly smaller discrete set of mass-radius curves are averaged over at a \textit{fixed} mass (see shaded region of the upper-left panel). The offset in radius of the posterior-averaged mass-radius curve (relative to the underlying dashed curve) never exceeds $4 \%$. In the lower-right panel this posterior-averaged radius offset reaches $\sim20 \%$ at masses $\sim2$ M$_{\odot}$. Note that although the solid blue lines in the left panels intersect the underlying mass-radius curves (dashed), the offset in the right panels does not go to zero; this is because the offsets are calculated as an average of absolute offsets, over a \textit{discrete} set of mass-radius curves.}
  \label{ExamplesScenario} 
\end{figure*}

In Fig. \ref{ExamplesScenario} we cast certain properties of the joint posterior distributions of the EOS parameters onto a mass-radius space, where again the upper- and lower-panels respectively correspond to the first and second scenarios. In the first scenario, the mass-radius curves within the $1\sigma$ credible region (left panels) are clustered around the underlying mass-radius curve at all masses. For the second scenario however, all mass-radius curves within the $1\sigma$ credible region exhibit drastic deviations from the underlying curve, corresponding to the EOS softening in the domain of the second polytrope (smaller pressures $P_2$ are favoured than that corresponding to the underlying EOS). 

In the right panels of Fig. \ref{ExamplesScenario} the average offsets in radius for (a subset of) mass-radius curves within the $1\sigma$ credible region is shown. In the first scenario the offset never exceeds $7 \%$. The shape of the curve can be explained via examination of the blue curve in the left panel: the difference between the average and the underlying mass-radius curve increases with mass until it reaches a local maximum at approximately $0.4$ M$_{\odot}$. The difference decreases, reaching a local minimum when the average and underlying mass-radius curve intersect, after which it increases again. The local minimum and sharp rise around $2.0$ M$_{\odot}$ and higher are artefacts of the method of calculating this measure. The local minimum: the mass-radius curves that are furthest from the underlying curve do not reach these masses and therefore do not contribute to the average offset anymore, causing the minimum (see also the caption). The sharp rise: at the maximum mass of the underlying mass-radius curve only the curves on the right side of the underlying curve contribute, shifting the average offset to higher values. In the second scenario the offset curve behaves in a more intuitive manner when compared to the posterior-averaged mass-radius curve in the corresponding right panel. For masses below $1.5$  M$_{\odot}$ the offset in radius is less than $3 \%$, but after this point there is a sharp increase to a maximum offset of almost $20 \%$ at $2.0$  M$_{\odot}$.

These two examples serve to illustrate the potential for distributional distortion if one is ignorant of pathological behaviour (such as Jacobian singularities) in a mapping between the spaces of exterior and interior parameters. The distortion is especially clear in the second scenario: the EOSs seemingly favoured by the data (and some prior) are appreciably softened in the domain of the second polytrope relative to the underlying EOS; we find the offset in radius to be reach as high as $20 \%$.

\subsection{Inclusion of radio pulsar mass constraints}
\label{sec:extramass}

\begin{figure*}
    \includegraphics[width=.9\columnwidth]{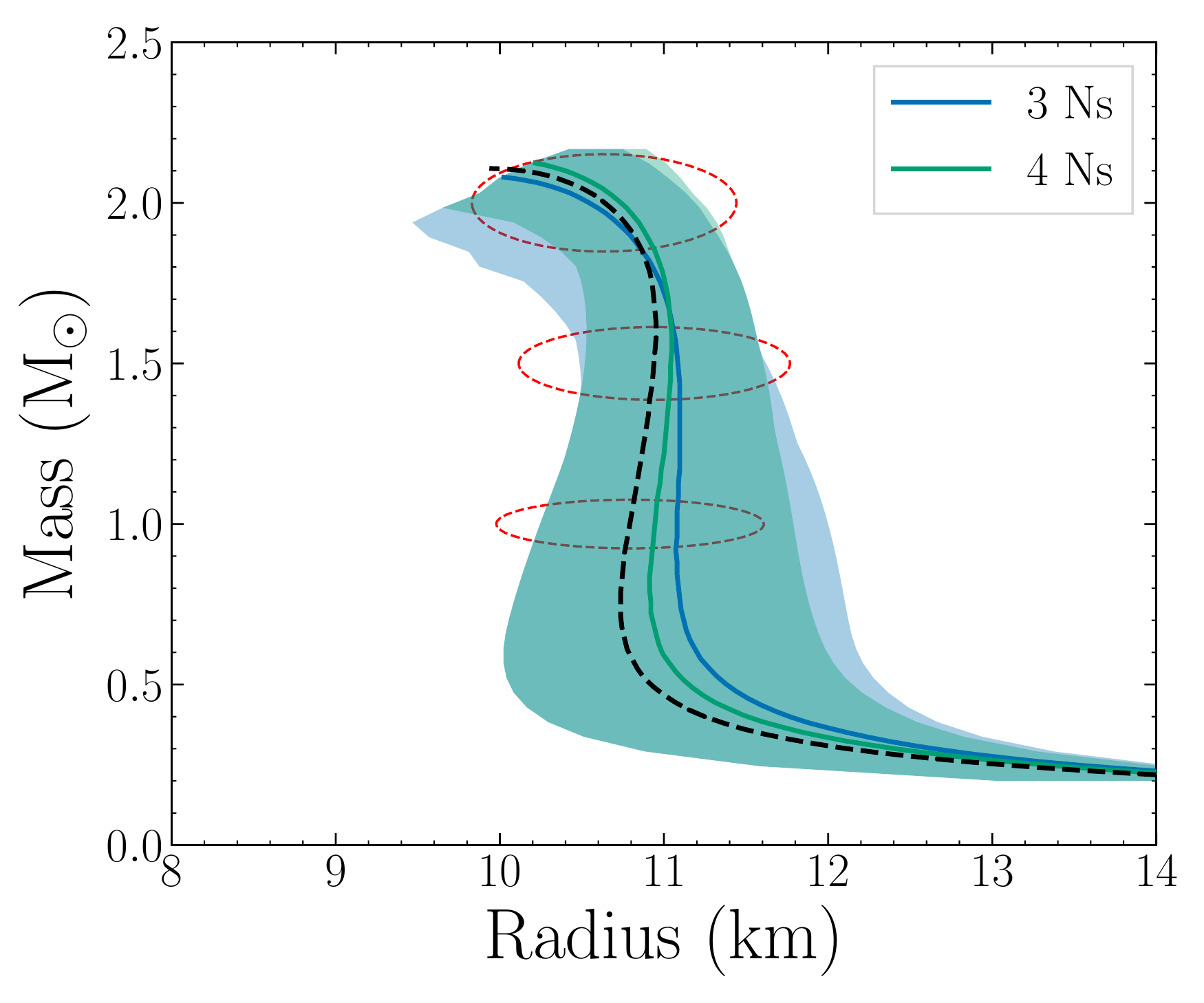} 
    \includegraphics[width=.9\columnwidth]{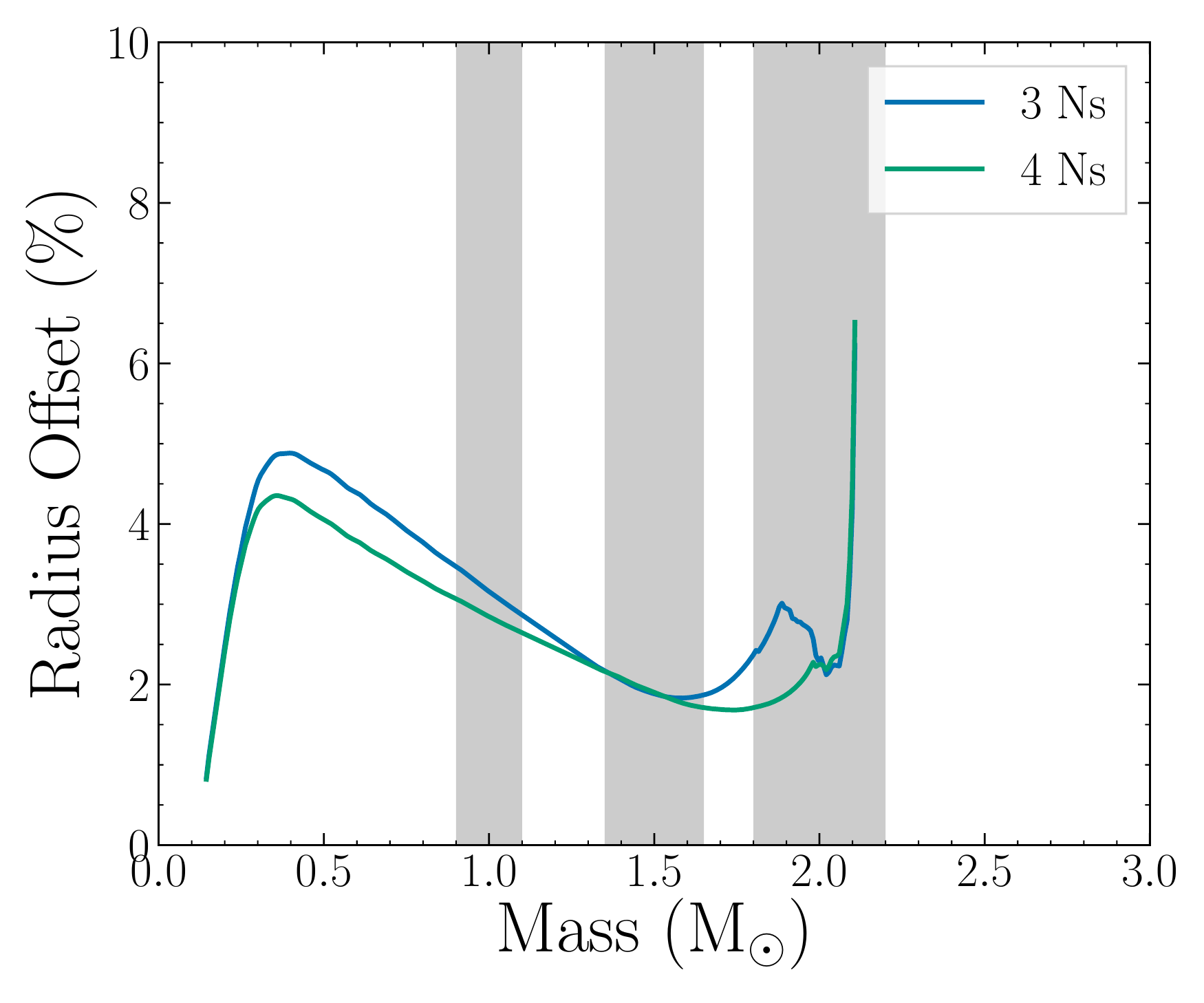} 
    \includegraphics[width=.9\columnwidth]{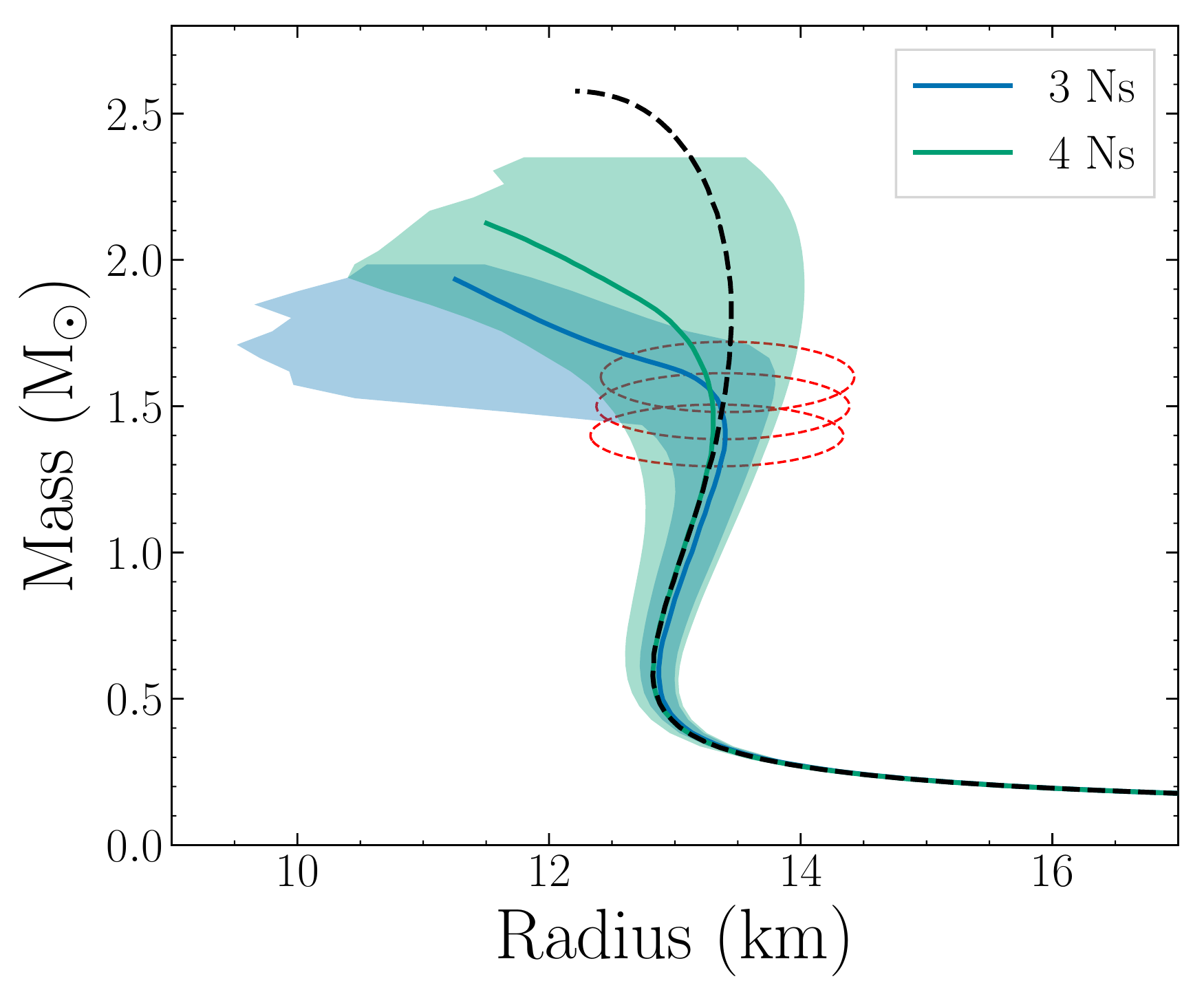}
    \includegraphics[width=.9\columnwidth]{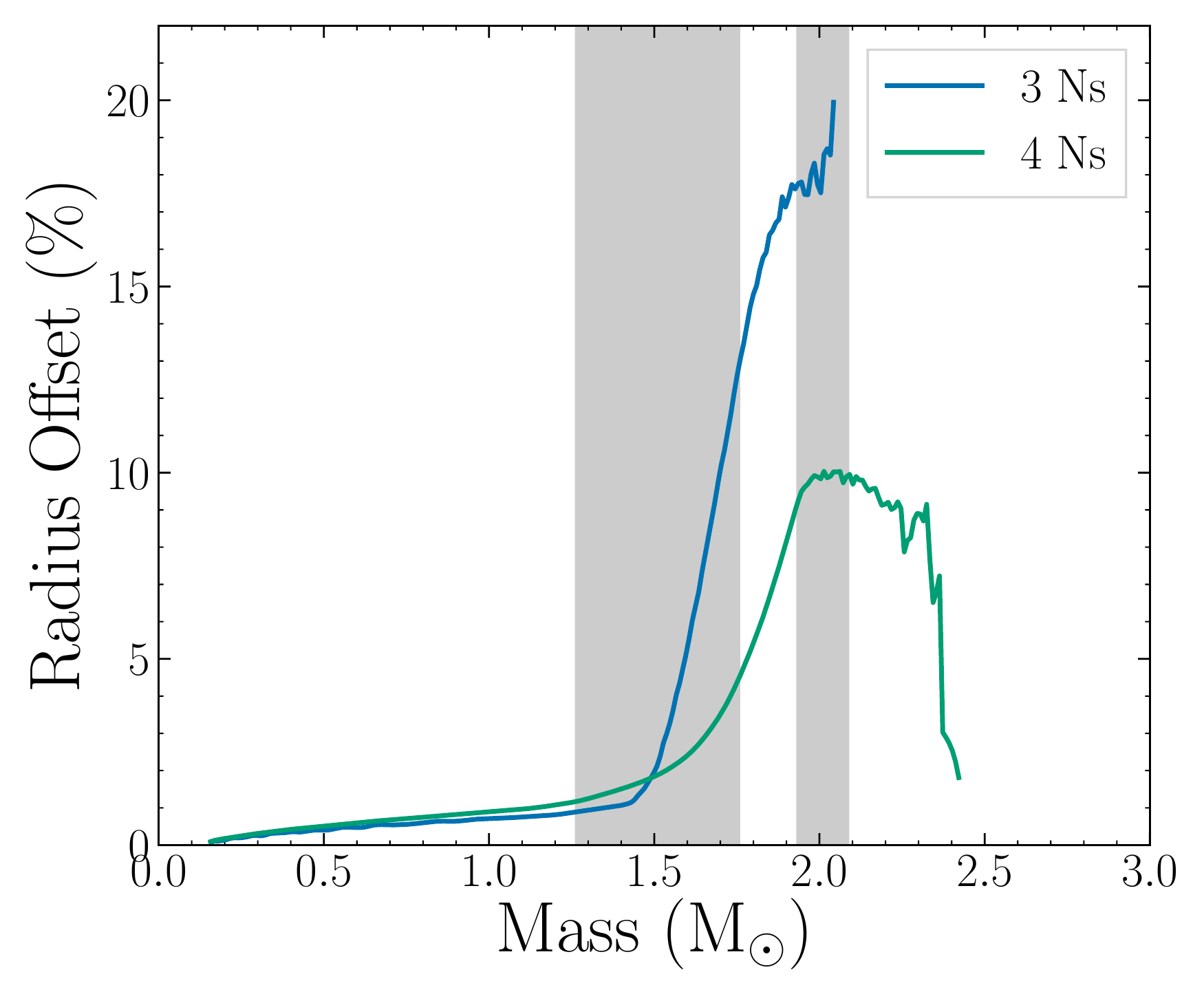}
  \caption{Similar to Fig. \ref{ExamplesScenario}, but with the inclusion of a one-dimensional Gaussian constituent posterior distribution approximating the \citet{Antoniadis13} pulsar radio timing mass constraint -- that is, the constraint with a modal mass of $2.01$ M$_{\odot}$. The effect in the upper panels is small: there already exists strong posterior for the underlying EOS due to the high-mass mass-radius posterior (red dashed ellipse). The average and the underlying mass-radius curves however more are comparable with the constraints from four stars. Moreover, the area spanned by mass-radius curves associated with EOSs within the $1\sigma$ credible region is narrower at higher masses (upper-left panel): the average offset in radius is significantly smaller (upper-right panel). In the lower panels, the additional mass constraint exerts a larger effect. The area spanned by mass-radius curves associated with EOSs within the $1\sigma$ credible region expands and shifts upwards (lower-left panel); the underlying EOS remains outside $1\sigma$ credible region, however. The average offset in radius for all masses is now less than $10 \%$ (lower-right panel).}
  \label{ExamplesScenario2M} 
\end{figure*}

In this section we discuss the statistical inclusion of existing radio pulsar mass constraints and demonstrate the change in the two example scenarios considered in Section \ref{sec:demonstration of distortion}. However, we also demonstrate that folding-in such constraints does not mitigate the problem of distributional distortion, but can result in distortion being less discernible upon inspection of the transformed distribution.

In recent years radio timing has produced a number of high-precision pulsar mass estimates \citep[for a review of the techniques used, see][]{Lorimer08}. The most interesting discoveries have been two massive radio pulsars PSR J$0348+0432$ by \citet{Antoniadis13} and PSR J$1614-2230$ by \citet{Demorest10} with masses of $2.01 \pm 0.04$ and $1.97 \pm 0.04$ M$_{\odot}$ respectively, although for the latter \citet{Fonseca16} have since reported a revised mass of $1.93 \pm 0.02$ M$_{\odot}$.

These mass constraints could be used as a form of prior on the parameters of an EOS model: e.g., any EOS that does not permit such massive stars could in practice be discarded (effectively equivalent \textit{a priori} assigned zero probability density), reducing the hypervolume of the EOS parameter space with finite prior support. An advantage of this modelling assumption is a boost in numerical resolution (e.g., a higher-resolution Jacobian library on parameter space) with fixed compute resources. However, to determine if the maximum mass permitted by a given EOS exceeds (approximately) two solar masses, the TOV equations need to be solved once for every EOS parameter vector (the maximum mass constraint cannot be imposed by disjointly cutting down the range of one or more of the EOS parameters). 
 
A more consistent manner in which to include the pulsar mass constraints is under the EOS parameter estimation framework described in Section \ref{sec:parameterest}: specifically, via redefinition of both the exterior parameter space, and the input joint posterior distribution of exterior spacetime parameters. The pulsar mass constraints can each be approximated as a one-dimensional Gaussian posterior distribution of a mass, given the first few reported moments \citep[expectation and variance; see, e.g.,][]{Antoniadis13} of a probability distribution \citep[assumed to be a likelihood distribution implicitly normalised with a uniform bounded mass prior; see, e.g.,][who adopt the reported moments as a likelihood approximation, but instead define a hierarchical mass prior]{Alsing17}.

The requirement that the dimensionalities of the parameter vectors $\bm{y}$ and $\bm{z}$ be equal is satisfied if both are redefined to include an additional mass parameter. For the general case, where total number of observed stars is greater than or equal to the number of EOS parameters ($s \geq n$), Equation (\ref{PosteriorM}) is separable into a product of: (i) a marginalisation over the product of a Jacobian determinant and a $2n$-dimensional joint posterior distribution of masses and radii; and (ii) a product of $s-n$ integrals, each over a one-dimensional pulsar posterior mass distribution (approximated from reported radio timing constraints as described above). Analytically we write
\begin{equation}
\label{eqn:fourthstar}
\begin{aligned}
\mathcal{P}(\bm{\theta} ~|~ \mathcal{D}, \mathcal{M}, \mathcal{I}) \propto & \mathop{\int} \left[\prod_{i=1}^{n}\mathcal{P} (M_i, R_i ~|~ \mathcal{D}_{i}, \mathcal{M}, \mathcal{I})\right]J(\bm{y})d^{n}\bm{M} \\
& \times \prod_{j=n+1}^{s} \int \mathcal{P} (M_j ~|~ \mathcal{D}_{j}, \mathcal{M}, \mathcal{I}) ~dM_{j}.
\end{aligned}
\end{equation}
For the case of one additional pulsar mass constraint (such that $s=n+1$), the joint posterior distribution of the EOS parameters is simply multiplied by a single integral, and the posterior distribution of the pulsar mass is given by
\begin{equation}
\mathcal{P} (M_{n+1} ~|~ \mathcal{D}_{n+1}, \mathcal{M}, \mathcal{I}) \propto \exp\left[- \frac{1}{2} \left(\frac{M_{n+1} - \widehat{M}_{n+1}}{\sigma_M} \right)^2\right].
\end{equation}

We note that incorporating pulsar mass estimates does not mitigate the problem of estimation of parameters of the (class of) piecewise-polytropic models. EOSs whose associated Jacobian determinant $J(\boldsymbol{y})$ is everywhere zero, will be assigned a posterior density of zero -- see Equation~(\ref{eqn:fourthstar}). By definition these EOSs are not finitely supported even if a pulsar mass estimate is incorporated. This is because the mapping remains non-invertible in astrophysical regions of the space of exterior parameters, where there may thus be non-negligible posterior support for exterior solutions. Whilst an extremely high pulsar mass estimate may shift the $1\sigma$ credible region towards the underlying EOS parameters, the posterior distribution is still severely distorted.

As an example, we show in Fig. \ref{ExamplesScenario2M} the effect of incorporating the $2.01$ M$_{\odot}$ pulsar \citep[as estimated by][]{Antoniadis13} on EOS inference for the two scenarios considered in Section \ref{sec:demonstration of distortion}. The change for the Case $3$ scenario is negligible, since the EOS parameters were already inferred from a constituent posterior with a modal mass higher than $2.01$. For the Case $2$ scenario we see that the change is much larger. The mass-radius curves within the $1\sigma$ credible region are now shifted more towards the underlying EOS in the left panel of Fig. \ref{ExamplesScenario2M}. The right panel shows that the average percentage offsets in radius with the inclusion of the pulsar decreases to below $10 \%$ for all masses. The underlying EOS parameters are however still not within the $1\sigma$ credible region of the distribution.

\section{Discussion}
\label{sec:discussion}
We conditioned on a piecewise-polytropic EOS model which does not have an associated invertible mapping between spaces of interior (source matter) parameters and exterior (spacetime) parameters. We then attempted to transform a set of distinct input joint posterior distributions of Schwarzschild gravitational masses and radii into joint posterior distributions of EOS parameters. The input distributions were chosen to span a large range of the astrophysical region of the space of exterior parameters.

As predicted on a theoretical basis, EOS parameter estimation is sensitive to the choice on which space the prior is defined. When the prior distribution is defined on the space of exterior parameters, distributional distortion is incurred because the Jacobian is singular in regions of parameter space -- that is, when no star has a central density in the domain of the third EOS polytrope (see Fig. \ref{fig:jacdeterminant} and Fig. \ref{fig:monotropes}). Despite lacking an exact metric for quantification of this distortion (because joint prior distributions cannot be meaningfully transformed between parameter spaces; refer to Section \ref{sec:parameterest}), we demonstrated that distortion is significant if the input joint posterior distribution supports exterior spacetime solutions similar to those permitted by stiffer EOSs when (upon input) there is no posterior support for high-mass stars (see Table \ref{tab1} and Fig. \ref{ProbEOS}) For such input distributions, the ``underlying'' EOS parameter vectors (see Section \ref{subsec:scenarios} for definition) are not within the $2\sigma$ credible region of the joint posterior distribution of EOS parameters. Moreover, the mass-radius curves corresponding to the EOS parameter vectors strongly supported following the transformation can be offset in radius by almost $20 \%$ in the most extreme cases (see Fig. \ref{ExamplesScenario}).

Folding-in existing constraints on the most massive known radio pulsar does not mitigate the problem (the mapping remains non-invertible in regions of the astrophysical exterior-parameter space where there may be non-negligible posterior support for exterior solutions), but can result in distortion being less discernible upon inspection of the transformed distribution (see Fig. \ref{ExamplesScenario2M}). In other words, when incorporating pulsar mass constraints the problem persists: this is clear from the fact that an underlying EOS parameter vector may not be recovered even within a $2\sigma$ credible region.

We proceed to discuss whether the posterior distortions demonstrated in this work are expected to be important for future EOS inference. The critical question to address is the following: is it plausible that the true (assumedly universal) EOS of matter constituting compact stars is (i) well-approximated by a piecewise-polytropic functional form, and (ii) exhibits an intermediate or high stiffness? There are several indicators that this is indeed the case.

The most compelling evidence derives from the tight constraints on the masses of several radio pulsars modal masses of $1.93$ M$_{\odot}$ and $2.01$ M$_{\odot}$). If these constraints are accurate and unbiased, EOSs which permit stable stars of masses below a maximum of $1.93$ M$_{\odot}$ are heavily disfavoured \textit{a posteriori} \citep[see, e.g.,][]{Alsing17}.

In recent years the \textit{radii} of several stars have been estimated via X-ray spectral modelling of moderately informative observational data sets \citep{Guver10, Steiner13, Ozel16, Nattila16, Nattila17}: in quiescence; after transient accretion episodes; or during/after accretion-driven thermonuclear bursts. These constraints typically favour radii in the range of $9 - 13$ km, although there exist a number of unresolved systematic uncertainties in X-ray spectral modelling, and the inferences of distinct research groups have exhibited contention \citep[see, e.g.,][]{Watts16}. It follows that at present these constraints have not proven as robust as those derived via radio pulsar timing.

The recent detection of gravitational waves generated by a binary neutron star merger spawned unique constraints on the EOS \citep{LIGO}: the (combined) signal was of a sufficiently high signal-to-noise ratio during late-inspiral for the likelihood to be sensitive to the (EOS dependent) binary tidal deformability entering in the waveform model. Although preliminary, the $90 \%$ credible region of the marginal joint posterior distribution of the tidal deformability parameters encompasses EOSs which permit non-rotating stars with radii as high 14 km. However, the most favoured EOSs permit stars with radii consistent with the constraints from X-ray spectral modelling of bursting, cooling, and quiescent stars. \citet{Annala17} show that combining the tidal deformability constraints with theoretical constraints from pQCD calculations results in a maximum radius of $13.4$ km for a $1.4$ M$_{\odot}$ compact star.

The EOS models consistent with the above observational constraints all lie in the region of function space where inference of comparable piecewise-polytropic functions is susceptible to severe inaccuracy, \textit{if} one attempts to transform a joint posterior distribution of exterior parameters onto the EOS parameter space (see Figure \ref{fig:monotropes}).

It is clear from our study that observing stars over a broad range in mass is helpful to minimise distortion. However, recent inference of the neutron star mass distribution suggests a peak in the range $1.3 - 1.5$ M$_{\odot}$ \citep{Ozel12,Alsing17}. Should the modal posterior mass-radius pairs for each observed star, obtained for example with future hard X-ray missions, fall within this range, the joint posterior distribution of EOS parameters will be distorted if the EP-paradigm methodology is applied. The pathological behaviour of the EOS parametrisation cannot be mitigated in any way by changing the definition of the set of exterior parameters.

We adopted analytical joint mass and radius posterior distributions loosely based (neglecting covariance and any departures from Gaussianity) on forecasted X-ray modelling constraints \citep[see, e.g.,][]{Miller15,Watts16}. However, the arguments presented in this paper regarding the risk of posterior distributional distortion due to an ill-defined prior apply whenever one first specifies a prior on a space of exterior parameters and then attempts an exterior-interior reparametrisation. We have demonstrated that piecewise-polytropic EOS parameter inferences are particularly sensitive to prior definition because the reparametrisation is non-invertible. The arguments are thus applicable to the use of pulsar moment of inertia constraints \citep[for slowly-rotating spacetimes; see][]{Kramer09}, and in gravitational wave astronomy \citep[see, e.g., the discussion on prior definition in][]{Lackey15}.

The EP-paradigm for EOS inference has appeared in the literature under the notion of probabilistic inversion given a deterministic mapping between interior source matter parameters and exterior spacetime parameters. However, this paradigm is ill-defined in general because said mappings are usually non-invertible. Although it may be possible to find a restrictive set of interior-exterior parameter mappings that are are at least everywhere locally injective, we advocate use of a different framework in order to avoid the types of issues raised in this work. A more principled approach which has also been applied in the literature is \textit{direct} EOS parameter estimation whereby a joint prior distribution is explicitly defined on a space of interior parameters (see, e.g., the IP-paradigm in Riley et al. submitted). In this case transformation of probability density between spaces of exterior and interior parameters is not required.

\section*{Acknowledgements}

The authors acknowledge support from ERC Starting Grant No. 639217 CSINEUTRONSTAR (PI Watts). We thank SURFsara (www.surfsara.nl) for the support in using the Lisa Compute Cluster. We are grateful to the referee for providing constructive comments.

\bibliographystyle{mnras}
\bibliography{references} 

\appendix

\begin{figure*}
     \includegraphics[width=.9\columnwidth]{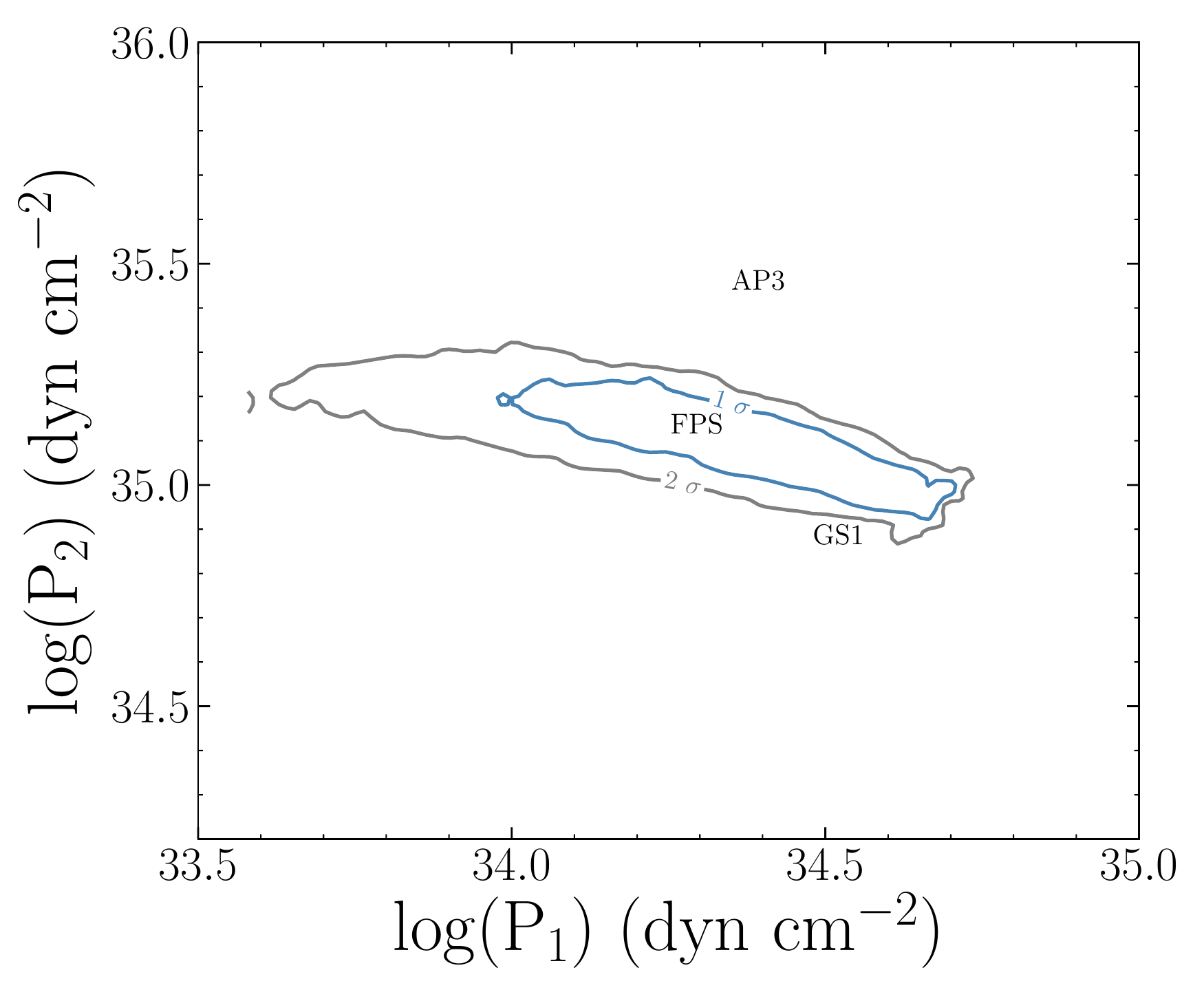}
    \includegraphics[width=.9\columnwidth]{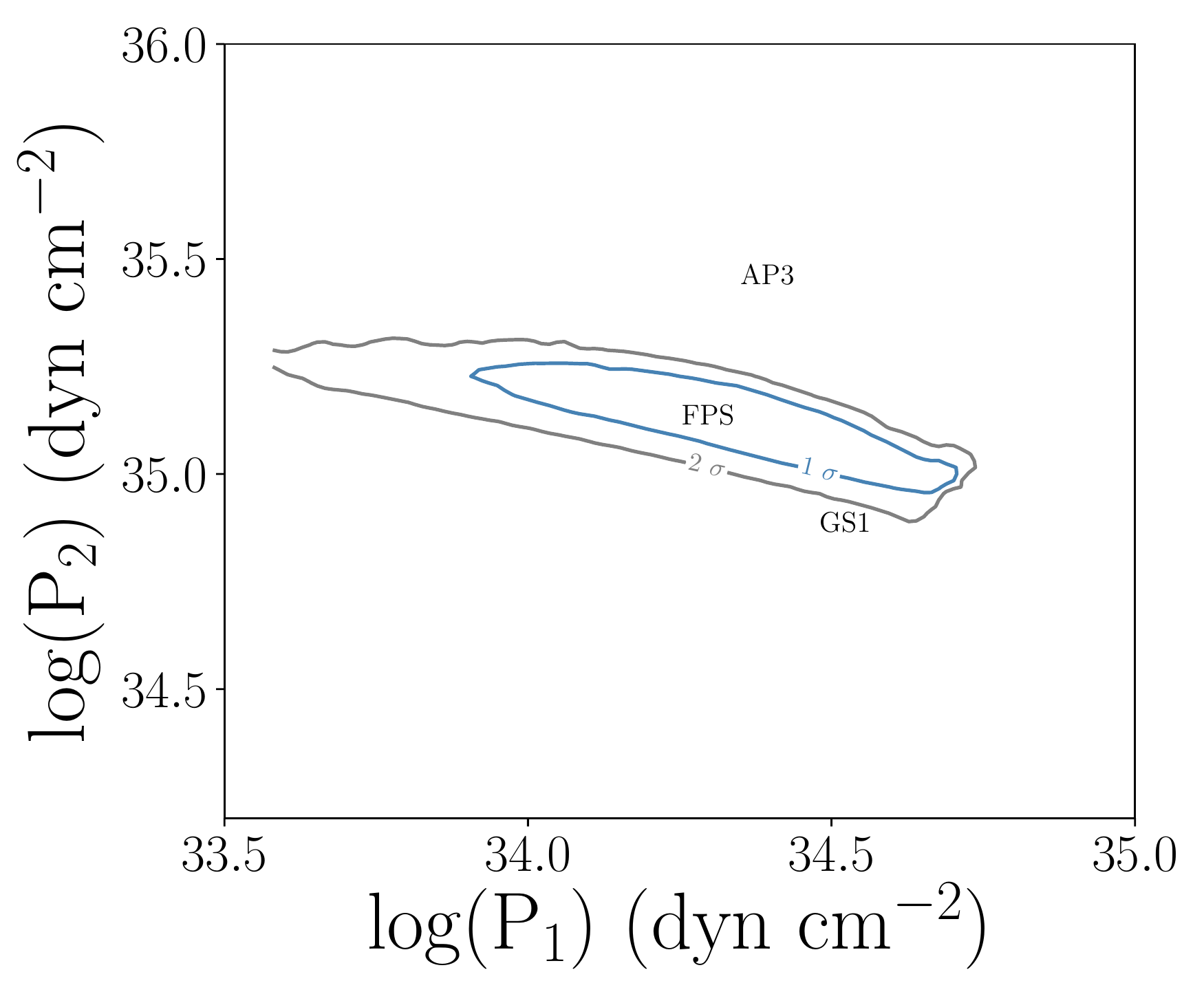}
    \includegraphics[width=.9\columnwidth]{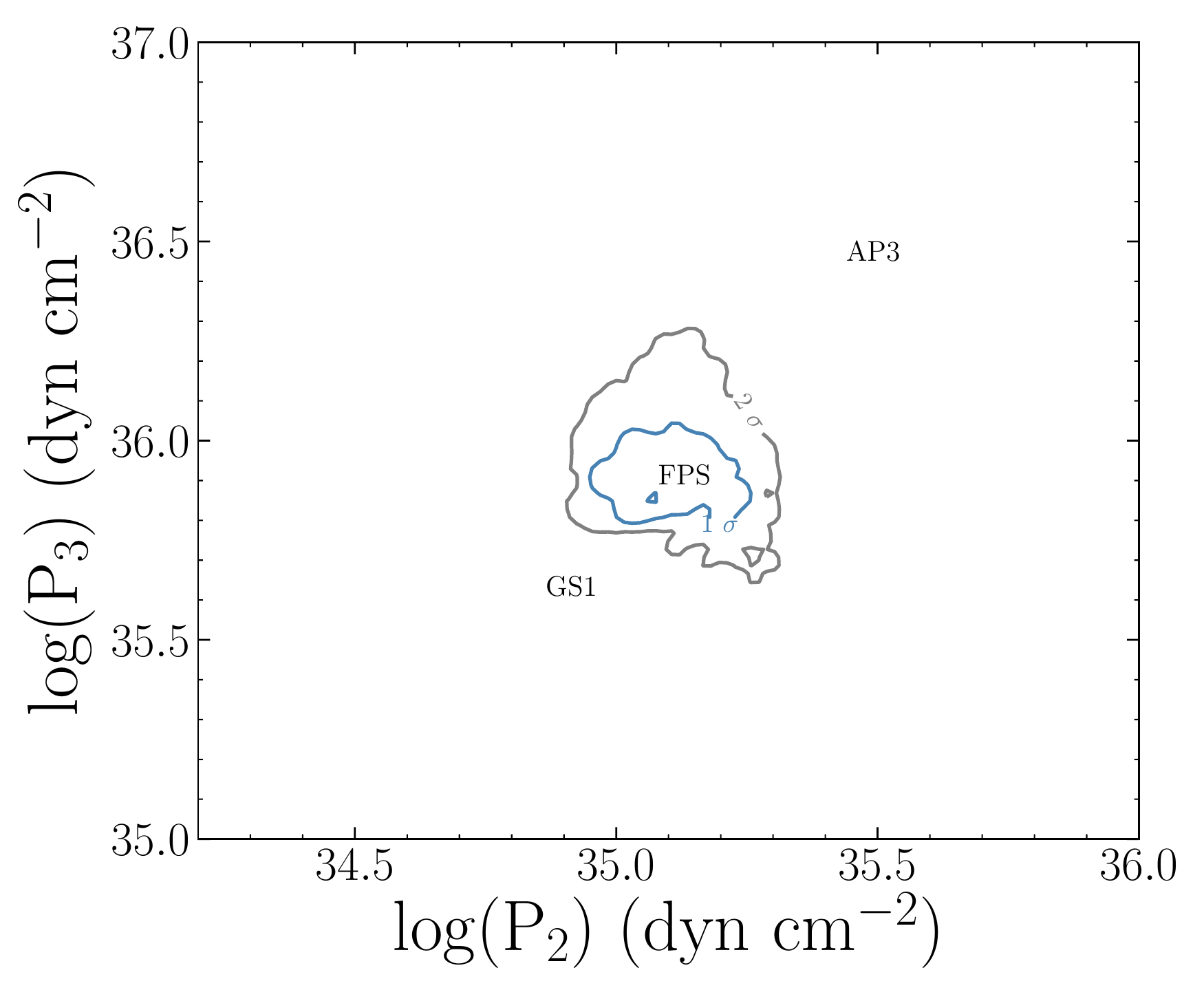}
    \includegraphics[width=.9\columnwidth]{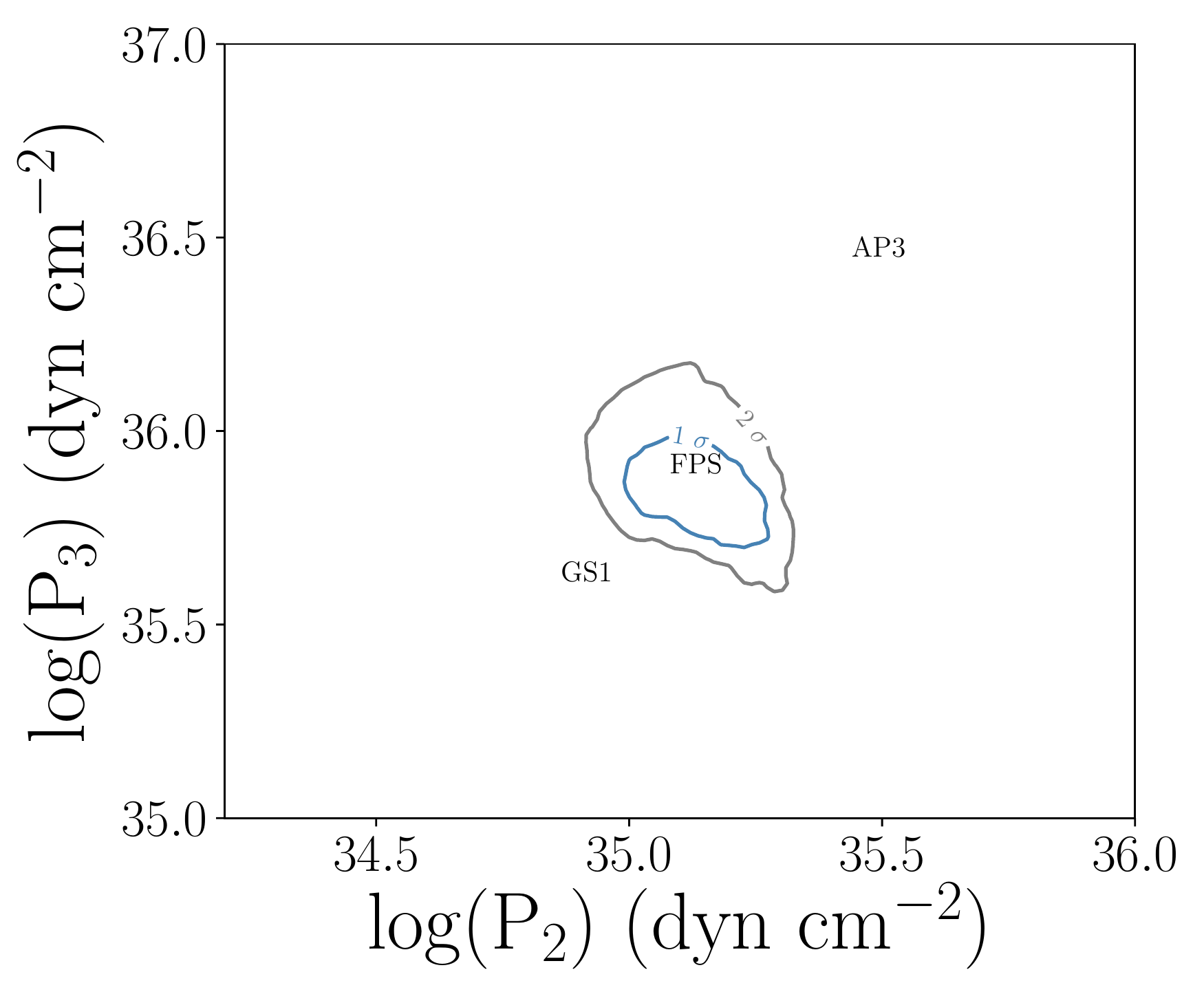}
  \caption{The numerical differences which manifest in the joint posterior distribution of the EOS parameters when using central densities (right) instead of masses (left) as the integration (marginalisation) variable -- see Equations~(\ref{PosteriorRho}) and (\ref{PosteriorM}) respectively. For this calculation the joint posterior distribution of masses and radii is defined as by a product of three mass-radius posterior distributions, each with a modal mass-radius pair which is permitted by the FPS EOS; this calculation is a reproduction of the calculation we display in Fig.~\ref{CompareOzel}, and is similar to a calculation performed by OP09. Remarkably, the forms of the posterior distributions based on the displayed (approximate) credible regions are close to congruent, but when central densities (the natural, interior parameters to use) are defined as the integration variables, the distribution is significantly smoother.}
  \label{ProbRho}
\end{figure*}

\section{Mass vs. central density}
\label{mvsrho}
In Section \ref{sec:rhotom} we have changed the marginalisation variables from central densities to masses. In this Appendix we demonstrate the numerical differences which arise between calculations with each type of marginalisation variable. 
We first note that, in order to change the integration variable from $\rho_c$ to $M$, it is required that for a given EOS, central density maps invertibly to gravitational mass. If this is true for all EOS parameter vectors considered (those with finite ``prior'' probability), masses are computationally advantageous as the marginalisation variables because the dimensionality of the Jacobian is reduced. For three stars, the the Jacobian reduces to a three-by-three matrix: the partial derivatives $\partial M_j/\partial \rho_c$, $\partial R_j/\partial \rho_c$, and $\partial M_j/\partial P_i$ are not required. Computational evaluation of a determinant is then faster, but it is only necessary to precompute a numerical Jacobian library on a (six-dimensional) parameter space. When integrating for a given EOS $\boldsymbol{\theta}$, determinants are approximated via three-dimensional (linear) interpolation (given the library), irrespective of whether masses or central densities are used. Computational cost is thus only reduced during the precomputation phase.

Central density is an interior parameter -- a boundary condition for integration of the TOV equations. It follows that it is natural to use central densities (that is, to transform to the interior parameter space, and not the joint space of EOS parameters and masses): the partial derivatives are numerically more accurate, and their dependence on the EOS parameters and gravitational mass is  smoother. As a result the final posterior distribution of the EOS parameters is smoother as well. In Fig. \ref{ProbRho} we show the difference when using masses in lieu of central densities for the example from OP09. It is clear that the general form of the transformed distributions are close to congruent, but when central densities are used, the $1\sigma$ and $2\sigma$ are appreciably smoother. 



\bsp	
\label{lastpage}
\end{document}